\documentclass[10pt,twocolumn]{article}   
\usepackage{geometry}                		
\usepackage{graphicx}				
\usepackage{authblk}
\usepackage{amsmath,amsfonts,amssymb,amsthm, bm}
\usepackage{indentfirst}
\usepackage{pifont}
\usepackage{soul}
\usepackage{xcolor}
\usepackage{hyperref}
\usepackage{abstract}
\usepackage{changepage}
\usepackage{anysize}
\marginsize{1.75cm}{1.75cm}{2cm}{2cm}
\let\OLDthebibliography\thebibliography
\renewcommand\thebibliography[1]{
  \OLDthebibliography{#1}
  \setlength{\parskip}{0pt}
  \setlength{\itemsep}{0pt plus 0.5ex}
}

\begin{document}

\twocolumn[{%

\begin{adjustwidth}{1cm}{1cm} 	

\begin{@twocolumnfalse}
  
\title{The impact of kinetic and global effects on ideal ballooning 2\textsuperscript{nd} stable pedestals of conventional and low aspect-ratio tokamaks}
\author[ ]{M.S. Anastopoulos Tzanis$^1$, M. Yang$^1$, A. Kleiner$^2$, J.F. Parisi$^{2,3}$,\\ G.M. Staebler$^1$ and P.B. Snyder$^{1,4}$}
\affil[1]{\it Oak Ridge National Laboratory, Oak Ridge, TN, USA}
\affil[2]{\it Princeton Plasma Physics Laboratory, Princeton University, Princeton, NJ, USA}
\affil[3]{\it Marathon Fusion, San Francisco, CA, USA}
\affil[4]{\it Commonwealth Fusion Systems, Devens, MA, USA}

\date{}	

\maketitle		

\begin{abstract}
The EPED model [P.B. Snyder {\it et al} 2011 {\it Nucl. Fusion} {\bf 51} 103016] has had success in describing the pedestal structure of Type-I ELM and QH-mode plasmas in tokamaks, by combining kinetic ballooning mode (KBM) and peeling-ballooning (PB) constraints. Within EPED, the KBM constraint is often approximated by a technique using calculated ideal ballooning mode (IBM) thresholds, fit to a functional form designed to capture non-local effects. It has been noted that quantitative differences between local ideal MHD and gyro-kinetic (GK) ballooning stability can be larger at low aspect ratio. KBM critical pedestals are consistent with observations in initial studies on conventional and spherical tokamaks. In this work, the application of a reduced model for the calculation of the kinetic ballooning stability boundary is presented based on a novel and newly developed Gyro-Fluid System (GFS) code [G.M. Staebler {\it et al} 2023 {\it Phys. Plasmas} {\bf 30} 102501]. GFS is observed to capture KBMs in DIII-D as well as the NSTX pedestals, opening a route to integrating this model into EPED. Finally, high but finite $n$ global ballooning modes are observed to limit the access to the local 2nd stability and thus provide a transport mechanism that constrains the pedestal evolution with $\beta_{p,ped}$. The high $n$ global ballooning stability is approximated by its ideal MHD analog using ELITE. It is shown that nearly-local high $n$ modes with $k_y\rho_s\sim0.25-0.5$ can provide a proxy for the critical $\beta_{p,ped}$ when a 2\textsuperscript{nd} stable access exists on DIII-D plasmas. The use of GFS and ELITE scaling in EPED provides improved agreement to EPED1 in an initial comparison with DIII-D pedestal data.
\end{abstract}

\vspace{0.5cm}

\end{@twocolumnfalse}

\end{adjustwidth}

}]

\section{Introduction}\label{sec1}

As efforts toward the design and construction of a tokamak Fusion Pilot Plant (FPP) intensify, the need for reliable predictive modeling capabilities becomes increasingly critical, as present-day tokamak experiments operate far from FPP-relevant parameter regimes. Meeting this need requires a deep physical understanding to develop reduced models that are both sufficiently accurate and computationally efficient for integration into FPP design and optimization workflows.

The overarching objective of such design loops is to achieve FPP configurations with adequate core performance to sustain high fusion power density. Tokamak fusion performance is commonly quantified using the triple product, $nT\tau_E$, where $n$ and $T$ denote the density and temperature of the deuterium--tritium plasma, and $\tau_E$ is the energy confinement time. In present devices, maximizing core pressure is often achieved through the formation of an edge transport barrier, resulting in the development of a pedestal upon which the core pressure is supported. Although this operational regime, known as H-mode, is highly beneficial for confinement, it is often accompanied by the destabilization of edge-localized modes (ELMs). In an FPP environment, even a small number of large ELMs  cause excessive sputtering, erosion, and melting of divertor components, making it essential to understand, predict, and ultimately control or avoid such events.

For the design of a tokamak FPP, pedestal conditions are therefore of central importance. Achieving significant fusion gain will likely require the plasma core to operate close to turbulence critical-gradient limits, rendering the pedestal height a key driver of overall performance. In addition, steady-state FPP operation relies heavily on non-inductive current drive, to which the pedestal contributes substantially through the bootstrap current. Accurate pedestal modeling is thus required to predict both performance and stability limits and to develop viable strategies for small or no-ELM operation with high bootstrap current fractions.

The most successful predictive model to date for pedestal structure and the onset of large type-I ELMs and QH-mode is the EPED model \cite{scaleDIIID,eped1}. EPED is based on two calculated constraints that govern the structure of the pedestal, characterized by its width, $\Delta_{\psi_N}$, and height, $\beta_{p,\mathrm{ped}}$. These quantities are related through a kinetic ballooning mode (KBM) constraint:
\begin{equation}
\Delta_{\psi_N} = c_1 \beta_{p,\mathrm{ped}}^{c_2},
\end{equation}
where
\begin{equation}
\beta_{p,\mathrm{ped}} = \frac{2 p_{\mathrm{ped}}}{\mu_0\left(I_p / L\right)^2},
\end{equation}
$p_{\mathrm{ped}}$ is the pedestal pressure, $I_p$ the plasma current, and $L$ the plasma boundary circumference. EPED incorporates a nearly-local ``stiff'' transport constraint based on KBM onset that limits the pedestal pressure gradient, together with a global constraint that limits the maximum pedestal width and height. As a result, the pedestal evolves close to the kinetic ballooning mode (KBM) limit, while large type-I ELMs or QH-mode are triggered when peeling--ballooning modes become unstable. The relationship between pedestal width and height therefore plays a central role in determining the onset of peeling--ballooning instabilities, making its accurate prediction essential.

In conventional aspect-ratio tokamaks, experiments find that $c_1 \sim 0.06$--$0.13$ and $c_2 \sim 0.5$ \cite{scaleDIIID,scaleJET,scaleCmod,scaleAUG,scaleJT60,scaleTCV}, consistent with theoretical predictions based on kinetic ballooning mode onset, often approximated via the ballooning critical pedestal technique using ideal ballooning calculations \cite{eped1}. A similar square-root scaling is observed in MAST, albeit with somewhat larger values of $c_1 \sim 0.12-0.15$ \cite{scaleMAST}. In contrast, NSTX experiments exhibit a markedly different scaling, with $c_1 \sim 0.4$ and $c_2 \sim 1$ \cite{scaleNSTX}. Ideal ballooning mode (IBM) stability calculations for NSTX reproduce an approximately linear dependence, $c_2 \sim 0.8-1$, but yield lower values of $c_1 \sim 0.2-0.3$, corresponding to significantly higher predicted $\beta_{p,\mathrm{ped}}$ than observed experimentally. This discrepancy has recently been attributed to differences between kinetic and ideal ballooning thresholds \cite{parisi1}, which appear to be enhanced in low aspect-ratio tokamaks. The inclusion of kinetic effects, along with an assumption that the onset of stiff transport occurs at the small values of the KBM growth rate, reduces the critical $\beta_{p,\mathrm{ped}}$, leading to a wider predicted pedestal.

An additional feature of both local ideal and kinetic ballooning stability is the existence of a 2\textsuperscript{nd} stable regime at large normalized pressure gradient $\alpha$, defined as
\begin{equation}
\alpha = -\frac{\mu_0}{2\pi^2}\frac{dV}{d\psi}\sqrt{\frac{V}{2\pi^2 R_0}}\frac{dp}{d\psi}
\end{equation}
where $V$ is the plasma volume and $R_0$ the major radius. For sufficiently low magnetic shear, $\hat{s} = d\ln q / d\ln r$, and shaped magnetic equilibria, no stability boundary exists with increasing $\alpha$. This behavior complicates the determination of pedestal width and height constraints based solely on local ballooning physics. Moreover, the applicability of local analyses in the pedestal remains an open question, as pedestal widths are often comparable to or smaller than the characteristic scale $k_y \rho_s$ of the most unstable KBMs. Global gyro-kinetic simulations have demonstrated that kinetic effects can close access to the 2\textsuperscript{nd} stable regime at low magnetic shear \cite{nonlocal}, although standard gyro-kinetic formulations remain intrinsically local. Perpendicular scale length of perturbations must be much smaller than equilibrium length scale variations. Recent work by Ref\cite{gbgk} introduce higher order terms to gyro-kinetics but no study had been performed to examine their impact in the pedestal. The role of finite $n$ effects in accessing second stability has also been examined within ideal MHD, highlighting the importance of global equilibrium variations through the interaction of high $n$ kink/peeling and ballooning modes \cite{localPB1,localPB2}.  In the BCP technique in the EPED model, non-local effects are approximated by a functional form that captures the local ideal limit in the first stable region, and projects downward following a theoretically predicted relationship between $d \beta_p/d \psi_N \sim \alpha$ and $\hat{s}$ meant to capture the closure of 2nd stable access by finite-$n$ modes.

Alternative predictive pedestal models such as IMEP \cite{imep1}, substitute the local KBM limit by observed empirical limits for the normalized average electron temperature gradient within the pedestal, where $R<\nabla T_e >/ T_{e,ped}=-82.5$ is evaluated from AUG, DIII-D, JET and C-MOD plasmas  \cite{imep2,imep3}. This limit sets the electron temperature profile, but the ion temperature and electron density also need to be specified. Within IMEP, this is achieved using the ASTRA transport solver \cite{astra} by defining the transport in the pedestal region. As a result, the electron particle diffusion and ion heat diffusion need to be specified. Those quantities are obtained considering neoclassical transport and an additional anomalous transport component, which is related to the electron heat diffusion and scaled based on empirical measurements. Although this approach provides a step in the right direction, it still relies on tuned transport coefficients for electron density and ion temperature which are tokamak specific hindering the extrapolation of those quantities for predictions of FPP designs. Such approaches would strongly benefit from a physics based predictive pedestal transport model. 

In this work, we calculate the coefficients in the KBM criticality relationship $\Delta_{\psi_N} = c_1 \beta_{p,\mathrm{ped}}^{c_2}$ using a newly developed Gyro-Fluid System (GFS) code \cite{GFS}, which has successfully captured KBM physics in NSTX. GFS solves a gyro-kinetic fluid moment system that retains key kinetic effects at a fraction of the computational cost of full gyro-kinetic simulations, enabling integration with frameworks such as EPED. In addition, the ideal MHD code ELITE \cite{ELITE} is employed to examine global effects on access to 2\textsuperscript{nd} stability, which are found to provide realistic width scaling for DIII-D plasmas. Although ELITE does not include kinetic physics, the computational expense of global gyro-kinetic simulations currently precludes their use in iterative predictive studies. By comparing pedestal constraints obtained from GFS and ELITE within EPED, improved agreement is achieved relative to EPED1.0 and EPED1.6 for a set of DIII-D plasmas.

This paper is organized as follows. Section\ref{sec2} reviews existing methodologies and introduces the GFS model, together with the procedure used to identify the local critical $\beta_{p,\mathrm{ped}}$ and determine the KBM criticality scaling. In additional, a methodology is provided for obtaining the KBM critiality scaling at 2\textsuperscript{nd} stability using ELITE to capture non-local effects which close off local 2\textsuperscript{nd} stability. Section\ref{sec3} presents predicted KBM criticality scalings for DIII-D and NSTX plasmas and examines the influence of geometry and plasma parameters on the local pedestal stability of NSTX \#139047. Section\ref{sec4} investigates the impact of global effects on 2\textsuperscript{nd} stable DIII-D pedestals using ideal ballooning stability and high but finite $n$ ballooning constraints. Section\ref{sec5} demonstrates the applicability of GFS and ELITE for computing KBM constraints within EPED for DIII-D pedestals. Finally, Section\ref{sec6} summarizes the results and discusses their implications.

\section{Creation of ballooning critical pedestals}\label{sec2}
 
As it is highlighted in Section\ref{sec1}, kinetic effects can be important for the determination of the critical height $\beta_{p,ped}$ for a fixed width $\Delta_{\psi_N}$. It is well known that ion drift resonance and trapped particle effects can further destabilize the kinetic analogue of the ideal ballooning mode close to marginal stability \cite{kinetic1}\cite{kinetic2}\cite{kinetic3}. Those effects are enhanced in low aspect ratio and strongly shaped plasmas, which creates the need for an accurate but fast model to capture those effects. The recent development of the GFS code enables a quick and accurate model to be coupled with integrated modeling tools. In the following two sub-sections, a review of the current understanding and methodology is presented for the {\it ``ballooning critical pedestal"} (BCP) \cite{eped1} and {\it ``gyro-kinetic critical pedestal"} (GCP) \cite{parisi3} methods to obtain the width/height evolution as well as a detailed description of the current approach using GFS is presented. Finally, alternative modeling considerations for 2\textsuperscript{nd} stable pedestals are also described in the following sections. 

\subsection{The KBM pedestal structure constraint}

The accuracy of the EPED model depends on the accurate calculation of constraints on the pedestal width and pedestal height. The relationship between $\Delta_{\psi_N}$ and $\beta_{p,\mathrm{ped}}$ determines the pedestal pressure gradient, $\nabla p$, which drives ballooning modes and, through its connection to the bootstrap current, also influences the kink/peeling drive. An accurate description of the coupled relationship between $\Delta_{\psi_N}$ and $\beta_{p,\mathrm{ped}}$ is therefore essential for understanding global peeling--ballooning stability and for obtaining reliable predictions of the pedestal structure.

From a physical perspective, this relationship is set by micro-instabilities during the inter-ELM cycle that are destabilized by (nearly) local kinetic gradients. The formation of the edge transport barrier during the H-mode transition leads to the stabilization of electrostatic turbulence through increases in $\beta_p$ and $\beta_p'$, where the prime denotes a radial derivative. This stabilization introduces a positive feedback mechanism that allows the pedestal to build up. However, when the pressure gradient becomes sufficiently large, KBMs are destabilized, resulting in enhanced local transport. This provides a limiting mechanism that prevents the local pressure gradient from exceeding a critical value. The resulting local critical gradient $\alpha_{crit}$, or equivalently $d \beta_p/d \psi_N$, governs the pedestal structure and its evolution as the pedestal forms.

A substantial body of literature from several tokamaks worldwide has examined the stability of local and global KBMs in the pedestal region, to which the reader is referred for further details (see, e.g.\ the non-exhaustive list in \cite{kbm1,kbm2,kbm3,kbm4,kbm5,kbm6,kbm7,kbm9,kbm10,kbm11,kbm12,kbm13}). These studies report mixed conclusions regarding whether unstable KBMs constitute the dominant or a subdominant instability. Nevertheless, even in cases where KBMs are found to be linearly stable, the associated $\alpha_{crit}$ is often observed to lie close to experimentally measured values. Furthermore, nonlinear gyro-kinetic and gyro-fluid simulations show that particle and heat fluxes increase sharply as $\alpha_{crit}$ is approached \cite{nlkbm1,nlkbm2,nlkbm3,nlkbm4,nlkbm5,nlkbm6}. These observations suggest that KBMs remain close to marginal stability throughout much of the pedestal region, making them a natural candidate for constraining pedestal evolution.

Limitations of this framework (given that KBMs are typically calculated in the local limit) can arise when a significant portion of the pedestal accesses local 2\textsuperscript{nd} stability. Under such conditions, other micro-instabilities may dominate transport as $\beta_p$($\beta'_p$) increases, most notably micro-tearing modes (MTMs), or trapped electron modes (TEMs) driven unstable by large density gradients. However, because different mechanisms may drive particle and heat transport in the ion and electron channels, identifying a robust transport mechanism in the absence of KBMs becomes challenging. In addition, global effects have been identified as a potential destabilizing mechanism within the 2\textsuperscript{nd} stable regime; further discussion of these effects is provided in a later section.

The simplicity of the KBM critical gradient, together with the associated stiff transport, provides a broadly applicable constraint for pedestal evolution and forms the basis of its successful implementation within the EPED model. In practice, however, enforcing a strictly local $\alpha_{crit}$ across the entire pedestal is not feasible, as local 2\textsuperscript{nd} stability can occur and such a prescription is not necessarily consistent with the global equilibrium. To address this limitation, the BCP technique was developed, approximating the KBM constraint using local ideal ballooning (IBM) calculations in the 1\textsuperscript{st}  stable region, and extrapolating using a functional form motivated by calculations of finite $n$ ballooning modes which close local 2\textsuperscript{nd} stability access. More recently, the GCP technique has been introduced as a gyro-kinetic extension of the BCP approach. These methods are discussed in the following sections.

\begin{itemize}
\item{\bf Ballooning critical pedestal model}
\end{itemize}

The BCP technique was developed to simplify the computationally expensive evaluation of KBM stability and to address the difficulty of enforcing a locally critical KBM gradient across the entire pedestal (particularly in the presence of regions which are locally 2\textsuperscript{nd} stable). Early studies demonstrated that the critical thresholds for KBMs and IBMs are not substantially different \cite{kbm4,kbm6,kbm12,nlkbm5}, although kinetic effects tend to slightly reduce the threshold \cite{kin1,kin2}. This is the case particularly if a finite growth rate threshold is established for the onset of stiff transport.  On this basis, the KBM constraint can be approximated using the IBM stability criterion. This approximation offers significant computational advantages, as the IBM stability problem reduces to a simple 1D eigenvalue calculation that can be solved very efficiently and yields a binary stability outcome. In contrast, gyro-kinetic calculations often exhibit a wide spectrum of instabilities as equilibrium parameters vary, making the identification of the KBM both challenging and computationally demanding.

Constructing a pedestal that is locally KBM (or IBM) critical would, in principle, require integrating the critical pressure gradient radially to determine the pedestal height. However, this approach necessitates the generation of self-consistent equilibria for prescribed profiles, including the associated bootstrap current. The resulting changes to the equilibrium modify the local magnetic shear $\hat{s}$ and, consequently, $\alpha_{crit}$ itself. As a result, enforcing locally critical profiles throughout the pedestal becomes impractical, if not impossible. 

To overcome this difficulty, and account approximately for non-local effects which close local 2\textsuperscript{nd} stability, the BCP method instead requires that the central half of the pedestal be either IBM unstable or in the 2\textsuperscript{nd} stability regime. Under these conditions, the method provides an estimate of the critical $\beta_{p,ped}$ even when a portion of the pedestal reside in 2\textsuperscript{nd} stability. It should be noted, however, that reliable predictions require at least part of the central pedestal region to be first ballooning unstable in order to anchor the calculation; when the entire central half lies in 2\textsuperscript{nd} stability, the BCP method is not applicable.

In the BCP approach, the equilibria constructed to assess IBM stability follow the same procedure as in EPED1. Global plasma parameters and the pedestal density (and separatrix density and temperature, if known) are treated as known inputs to the model. The pedestal temperature, and hence the pressure, is then increased until the central half of the pedestal becomes BCP critical, that is, either IBM unstable or 2\textsuperscript{nd} stable. The resulting equilibria are bootstrap-consistent, which is essential for accurately capturing modifications to $\hat{s}$ and their impact on $\alpha_{crit}$. Within EPED, the bootstrap current is calculated using the Sauter model.  The BCP technique fits the ballooning criticality calculations to a functional form that is designed to accurately capture the ballooning limit in the 1\textsuperscript{st}  stable region, while approximating the effect of non-local modes closing 2\textsuperscript{nd}  stability access in the locally 2\textsuperscript{nd} stable region. In predictive calculations with EPED1.6, the BCP technique is typically applied with equal temperature profiles for all kinetic species and identical pedestal widths for the density and temperature profiles, though this is not a requirement of the technique itself, and it has been used, for example, to explore the impact of shifts of the density profile relative to the temperature profile.  

\begin{itemize}
\item{\bf Gyro-kinetic critical pedestal model}
\end{itemize}

As mentioned earlier, the BCP technique employs IBM stability as a proxy for KBMs. While this approximation has been shown to be adequate for conventional aspect-ratio tokamaks, deviations are observed in spherical tokamaks, where kinetic effects become more significant. To account for these effects, the GCP technique was developed, aiming to evaluate KBM stability within the central half of the pedestal. The equilibrium reconstruction as the pedestal pressure is increased follows the same procedure as in the BCP framework. However, because different equilibrium codes are employed, small differences in the calculated current density are expected, due to differences in the prescribed inductive current and bootstrap current model.

The principal distinction between the BCP and GCP techniques lies in the identification of KBMs. Owing to the wide range of instabilities present in gyro-kinetic calculations, a {\it ``fingerprint''} technique \cite{finger} is employed to identify KBM-like modes. Specifically, the ratio of heat to particle flux is used to discriminate between ion/electron temperature gradient modes (ITGs/ETGs), TEMs, and KBM instabilities, while mode parity is used to identify MTMs. In addition, due to their similar transport behavior, KBMs can be distinguished from TEMs by varying $\beta$ and examining the sign of $\mathrm{d}\gamma/\mathrm{d}\beta$. Finally, gyro-kinetic instabilities depend on the binormal wavenumber $k_y \rho_s$, and in the GCP framework $k_y \rho_s<0.2$ is typically considered. 

An important difference of the GCP approach, relative to BCP, is the absence of a direct mechanism to account for access to the 2\textsuperscript{nd} stability regime. Instead, the GCP framework identifies two distinct regimes corresponding to a wide branch (associated with the 1\textsuperscript{st} unstable boundary) and a narrow branch (associated with the 2\textsuperscript{nd} stable boundary). It is worth noting that, similarly to local IBMs, KBMs can also be stabilized at sufficiently low $\hat{s}$, indicating that access to 2\textsuperscript{nd} stability remains possible within a kinetic description. 

\subsection{Locally critical $\beta_{p,\mathrm{ped}}$ pedestal model for the width scaling}

In this section, a detailed description for the procedure used to calculate local critical $\beta_{p,\mathrm{ped}}$ using GFS is given. This information is then used to evaluate the KBM criticality scaling. The main difference between this approach and the BCP or GCP technique lies on the evaluation of locally critical heights $\beta_{p}$ as a function of local widths $\Delta$, rather than requiring the central half of the pedestal to be BCP/GCP critical. This approach attempts to provide a closer connection with integrating local critical gradients, $\mathrm{d}\beta_p/\mathrm{d}\Delta_{\psi_N}$, in order to obtain the relation between pedestal width and height. 

\begin{itemize}
\item{\bf The GFS code and KBM identification}
\end{itemize}

The GFS code solves the fluid (velocity) moments of the linear gyro-kinetic equation using an orthogonal set of basis functions defined along the field line and in velocity space, thereby reducing the computational complexity of the system. The full electromagnetic formulation is retained, including perturbations to the electrostatic potential, $\delta\phi$, the parallel component of the vector potential, $\delta A_\parallel$, and the parallel magnetic field, $\delta B_\parallel$. These fields are coupled self-consistently through quasineutrality and Amp\`ere’s law. Pitch-angle scattering and like-species collisions are included via a linearized collision operator. The space-velocity representation employs Hermite polynomials in the parallel spatial and parallel energy coordinates, and Laguerre polynomials in the perpendicular velocity coordinate. A detailed description of the GFS formulation is provided in Ref\cite{GFS}.
\begin{figure}[t!]
\centering
a) \includegraphics[height=5cm]{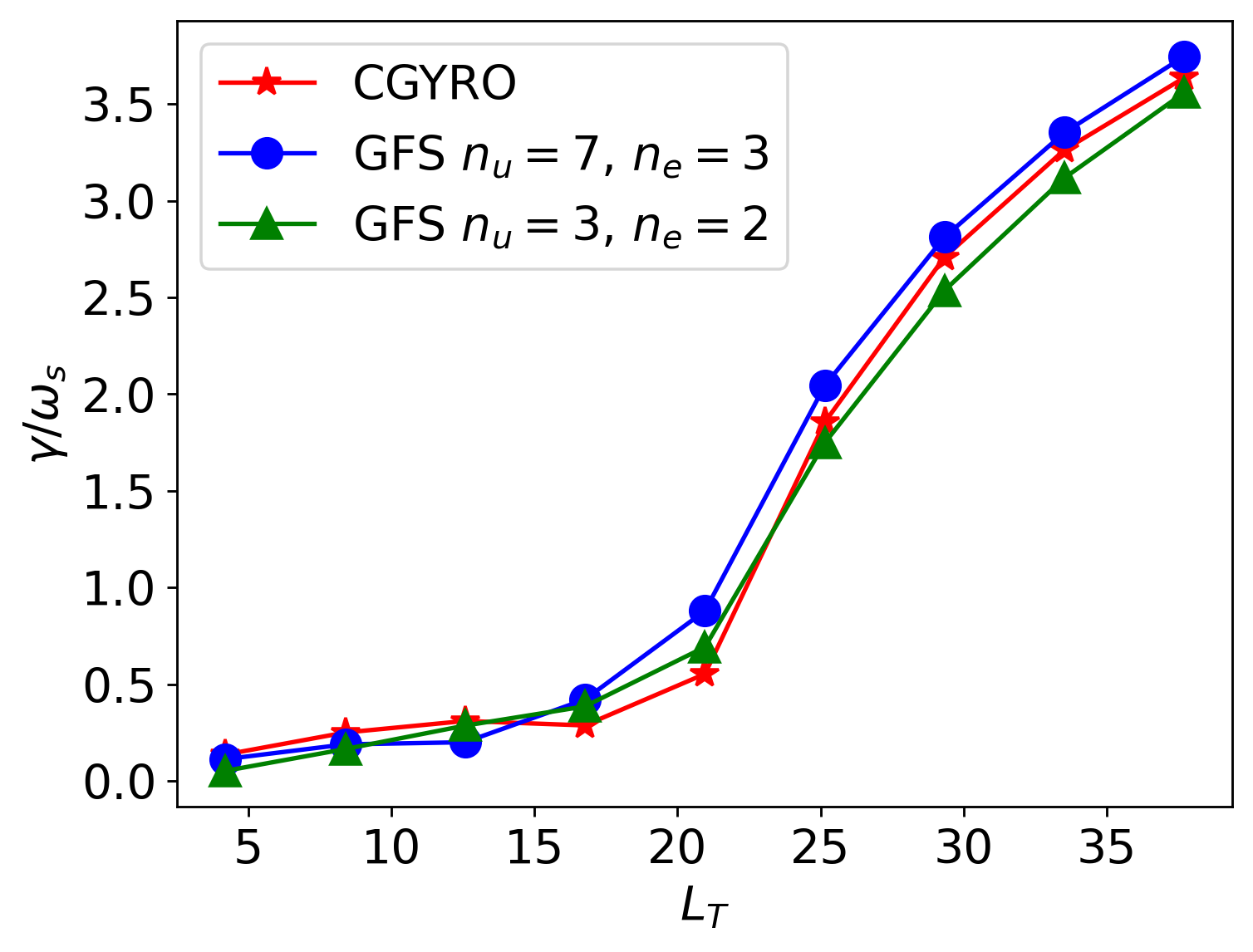} \\ b) \includegraphics[height=5cm]{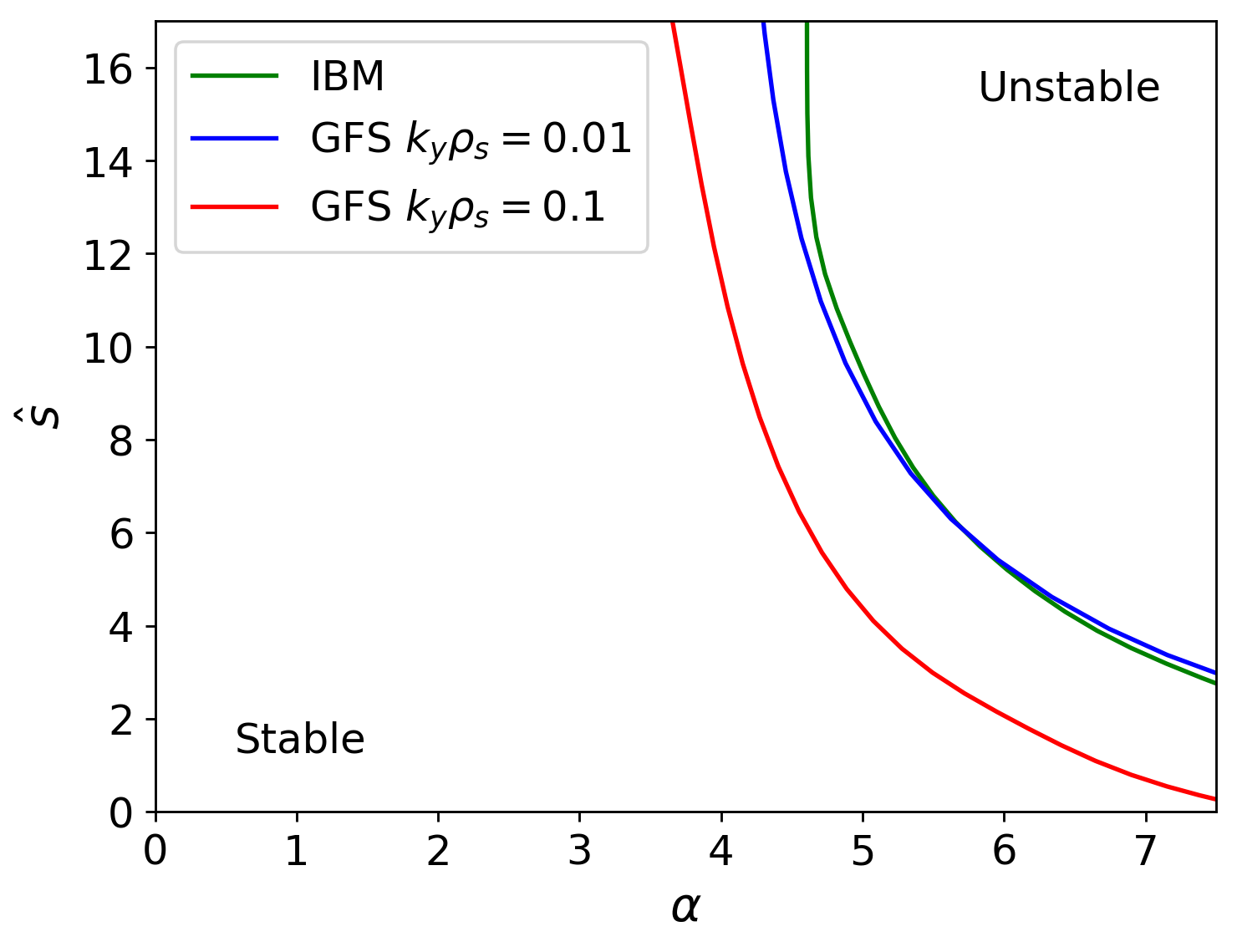}
\caption{a) Comparison of normalized growth rate $\gamma/\omega_s$ between GFS and CGYRO with varying temperature gradient normalized length scale $L_T$ for $k_y\rho_s=0.1$. b) The $\hat{s}-\alpha$ diagram for the IBM stability boundary and KBM stability boundary as calculated from GFS.}
\label{fig1}
\end{figure}

The capability of GFS to capture KBM physics has been demonstrated through extensive benchmarking against CGYRO \cite{cgyro} for a large database of NSTX pedestal cases ($\sim 800$ simulations). As reported in Refs\cite{myang,kinsey}, GFS reproduces both the linear frequencies and growth rates of KBMs with good accuracy, even at relatively low velocity resolution. The KBM threshold is identified by determining the value of $\beta$ or $L_n$, $L_T$ and $L_p$ at which the dominant instability transitions from electrostatic to electromagnetic. $L_n$, $L_T$ and $L_p$ are the density, temperature and pressure gradient normalized scale lengths respectively. Owing to the strong stiffness of KBMs with respect to $\beta$ or the kinetic gradients, a sharp increase in the growth rate is observed signaling the onset of KBMs. 

An example of this behavior is shown in Fig.\ref{fig1}, which compares GFS and CGYRO results. Good agreement is obtained for the normalized growth rate, $\gamma/\omega_s$, even at low parallel ($n_u$) and perpendicular ($n_e$) velocity resolution, where $\omega_s = c_s/a$ and $c_s$ is the sound speed. The use of low $k_y\rho_s \sim 0.05$--$0.1$ suppresses ITGs/ETGs modes as well as TEMs, while the reduced velocity resolution is sufficient to capture the critical KBM behavior and tends to exclude MTMs. In this manner, the KBM onset can be robustly identified within the GFS framework. In this example, $L_T$ is increased ($\beta'$ is changed consistently) until the KBM is destabilized at $L_T\sim20$.   

In addition, GFS is used to compute the local $\hat{s}$--$\alpha$ KBM stability boundary and to compare it with the corresponding IBM boundary. As shown in Fig.\ref{fig1}, in the limit $k_y\rho_s \rightarrow 0$ (equivalently $\omega_p \rightarrow 0$), the KBM reduces to the IBM, where $k_y$ is the binormal wavenumber, $\rho_s$ is the ion sound Larmor radius, and $\omega_p$ is the diamagnetic drift frequency. When finite $k_y\rho_s$ effects are included, the stability boundary shifts to lower values of $\alpha$, reflecting the additional destabilization of KBMs by ion drift resonance.

\begin{itemize}
\item{\bf Calculating $\bm{ \Delta_{\psi_N}=c_1\beta_{p,\mathrm{ped}}^{c_2} }$ scaling}
\end{itemize}

The next step is the calculation of the predicted pedestal width scaling for a set of consistent equilibria. This requires the specification of a set of basic plasma parameters, including the toroidal magnetic field $B_T$, plasma current $I_p$, normalized beta $\beta_N$ (this can be determined self-consistently in coupled core-pedestal calculations, but must be specified when the pedestal model is run in isolation), and effective charge $Z_{\mathrm{eff}}$, as well as the plasma boundary shape parameters $(R, a, \kappa, \delta, \zeta)$ and the electron density characteristics $(n_{e,\mathrm{ped}}, n_{e,\mathrm{sep}}, n_{e,0})$, which are treated as fixed inputs. The electron temperature at the separatrix, $T_{e,\mathrm{sep}}$, is also prescribed (separatrix quantities can be self-consistently calculated via integrated pedestal-SOL modeling, but must be specified as boundary conditions when running the pedestal model in isolation). The core electron temperature, $T_{e,0}$, is adjusted to achieve the target normalized $\beta_N = \langle \beta \rangle {a B_0}/{I_p}$, while the pedestal electron temperature, $T_{e,\mathrm{ped}}$, is varied. Here $\langle \beta \rangle = 2\mu_0 \langle p \rangle / B_0^2$, with $\langle p \rangle$ denoting the volume-averaged pressure and $B_0$ the on-axis toroidal magnetic field.

A sequence of equilibria is generated and iterated in this procedure, including a self-consistent bootstrap current calculated using the Sauter model \cite{sauter}. For cases of higher collisionality, the more accurate analytical bootstrap current formula of the Redl model \cite{redl} is also available and implemented within the equilibrium solver. The kinetic density and temperature profiles are represented using a \textit{tanh} function in the pedestal region and a polynomial form, $(1-\psi_N^\alpha)^\beta$, in the plasma core. In the present analysis, the pedestal widths of density and temperature are assumed to be equal, $\Delta_n = \Delta_T$, the ion and electron temperatures are taken to be equal, $T_e = T_s$, and the normalized density profiles are identical, $n_e/n_{e,0} = n_s/n_{s,0}$. The target toroidal plasma current $I_p$ is obtained by supplementing the bootstrap current with an inductive toroidal current density component, also described by a polynomial profile, which is adjusted to ensure that the core safety factor satisfies $q_0 > 1$, following the same procedure as in EPED.
\begin{figure}[t!]
\centering
a) \includegraphics[height=5cm]{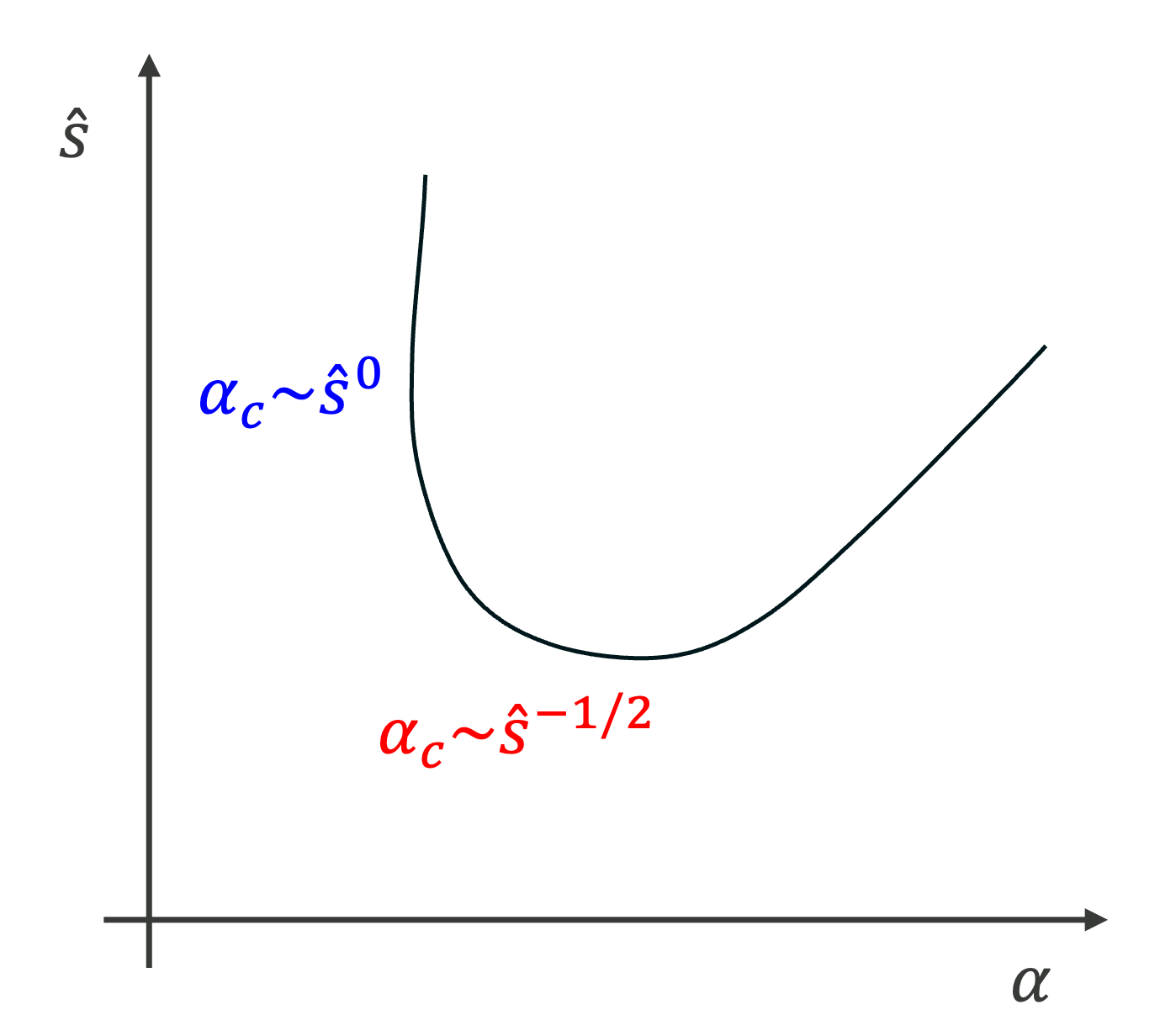} \\ b) \includegraphics[height=5cm]{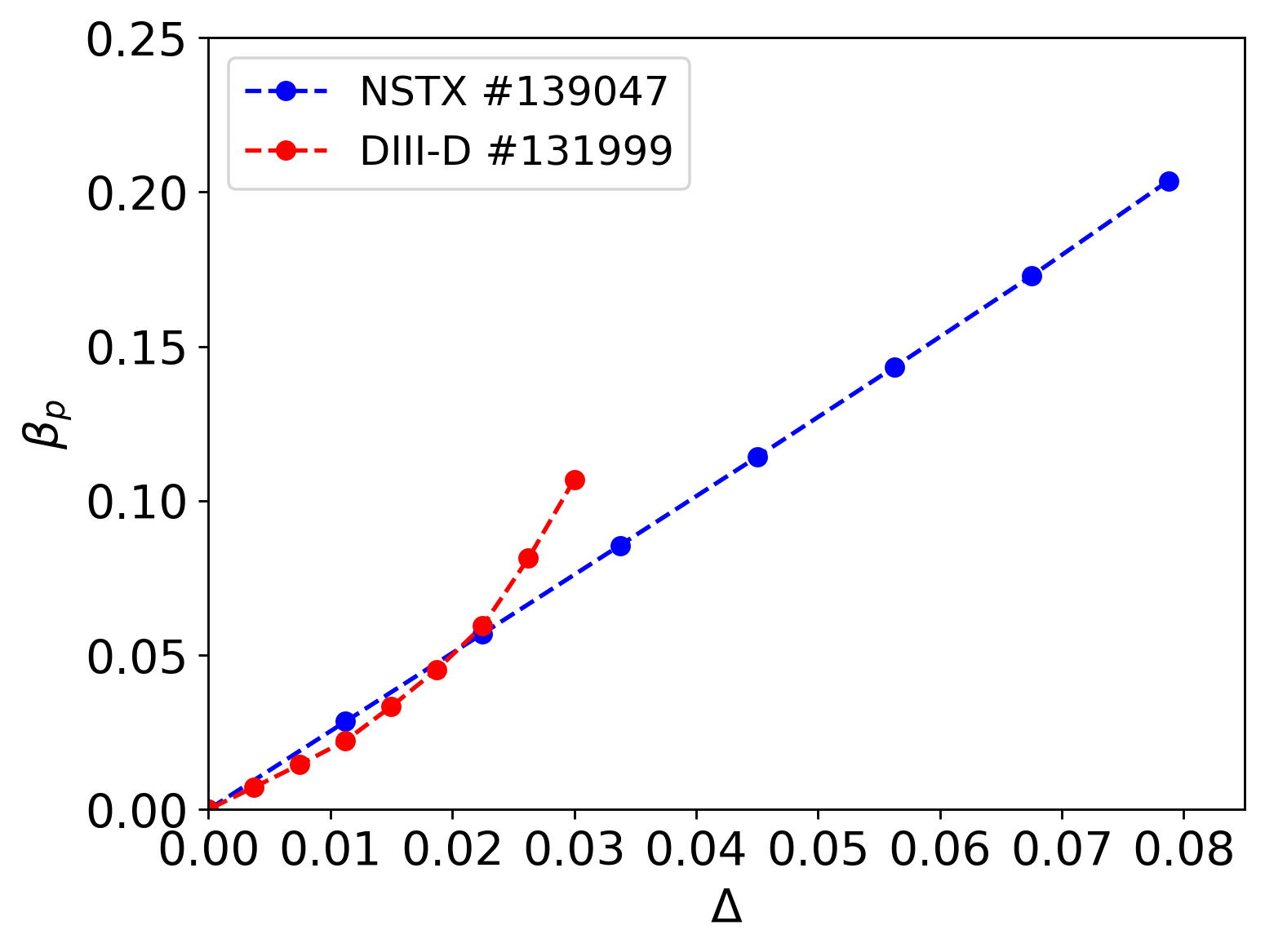}
\caption{a) Schematic of a typical $\hat{s}-\alpha$ diagram and the parametric dependence of $\alpha$ with $\hat{s}$. The black line (-) correspond to the ballooning boundary. b) The local critical height $\beta_p$ as a function of the local width $\Delta$ for DIII-D \#131999 and NSTX \#139047.}
\label{sa}
\end{figure}
\begin{figure}[t!]
\centering
\includegraphics[height=5cm]{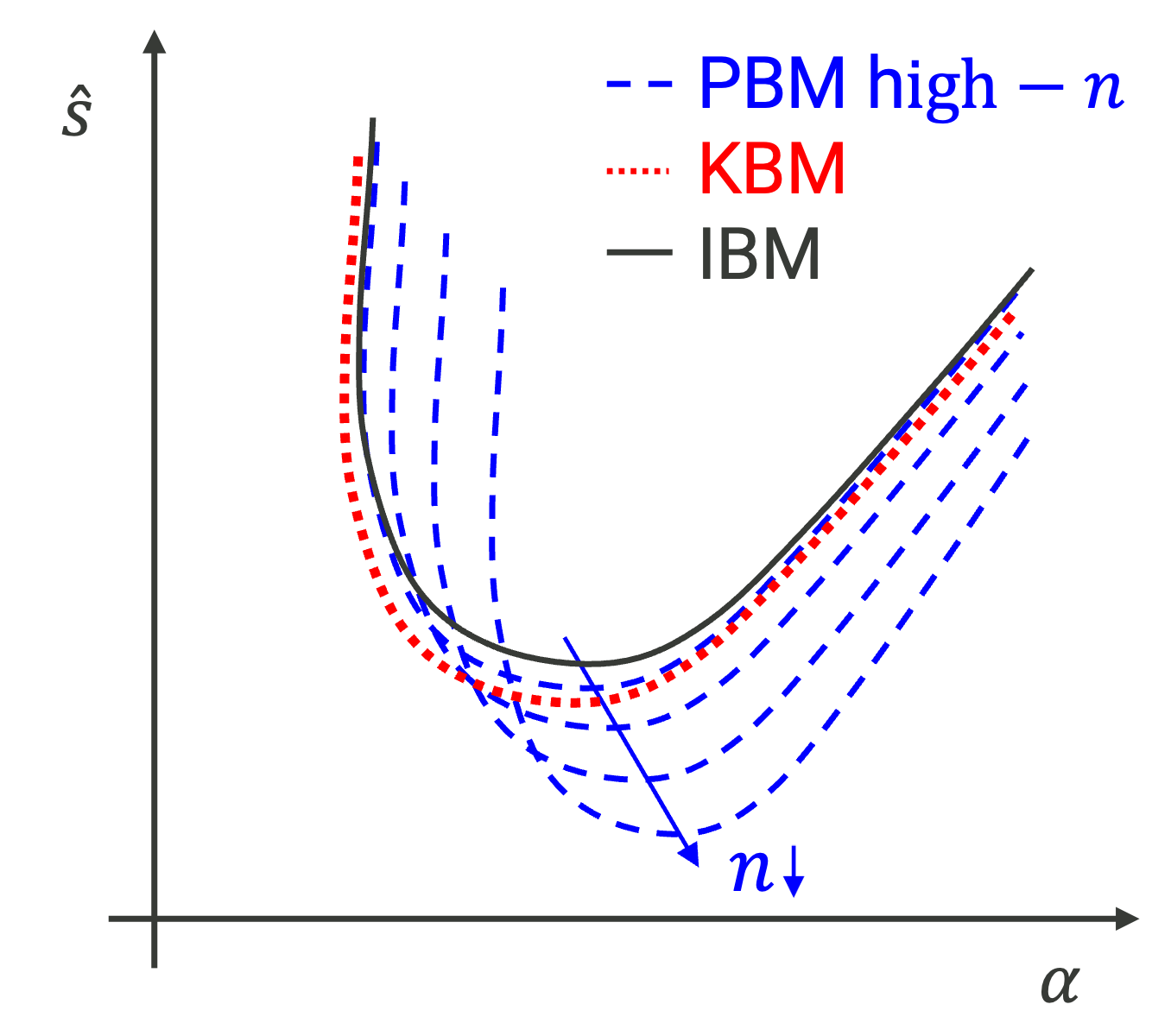} 
\caption{Schematic of the ballooning boundary in the $\hat{s}-\alpha$ diagram and comparison between local ideal (IBM), kinetic (KBM), and global high-$n$ peeling-ballooning (PBM) mode. The arrow indicates the direction of decreasing $n$. The peeling component of the global stability boundary is not depicted in this schematic picture.}
\label{glob}
\end{figure}

The local KBM threshold is identified through local scans in either $\beta$ or the normalized gradient scale lengths $L_n$, $L_T$, and $L_p$. In the remainder of this work, scans in $\beta$, with a consistent adjustment of $\beta'$, are employed. For a given equilibrium, the framework evaluates whether the equilibrium value of local $\beta$ matches the local KBM critical value, $\beta_{\mathrm{crit}}$, and adjusts the pedestal electron temperature, $T_{e,\mathrm{ped}}$, until consistency is achieved. As discussed earlier, the ultimate objective is to relate the local width in normalized poloidal flux, $\Delta_{\psi_N}$ (hereafter we drop the $\psi_N$ subscript), to the local height, $\beta_{p}$, by effectively integrating the local critical gradient $\alpha_{crit}$ across the transport barrier. In essence, we are interested in characterizing $\alpha\propto\mathrm{d}\beta_p/\mathrm{d}\Delta$, where $\Delta$ and $\beta_p$ are the local widths and heights of a given pedestal.

To illustrate this process, consider a pedestal structure characterized by a width $\Delta_{\psi_N}$ and height $\beta_{p,\mathrm{ped}}$. Within the pedestal, local values of the width, $\Delta$, and height, $\beta_p$, can be defined. Near the pedestal edge, where $\hat{s}$ is large, KBMs are readily destabilized as $\beta_p$ increases, corresponding to a relatively low critical gradient $\alpha_{crit}$. Progressing inward across the pedestal, $\hat{s}$ decreases and progressively larger values of $\beta_p$ can be sustained, because the local $\alpha_{crit}$ increases as the pedestal approaches to 2\textsuperscript{nd} stability. Since $\mathrm{d}\alpha_{crit}/\mathrm{d}\Delta > 0$ in this case, the resulting scaling yields a coefficient $c_2 < 1$. In contrast, in cases where $\hat{s}$ remains large throughout the pedestal and $\alpha_{crit}$ is approximately constant, the resulting scaling approaches $c_2 \sim 1$. 

A schematic $\hat{s}$–$\alpha$ diagram is shown in Fig.\ref{sa}, illustrating that at high $\hat{s}$ the critical gradient $\alpha_{crit}$ is nearly constant. The same figure also shows the local dependence of $\mathrm{d}\beta_p/\mathrm{d}\Delta \sim \beta_p/\Delta$ for DIII-D \#131999 and NSTX \#139047. For DIII-D \#131999, $\beta_p$ increases approximately quadratically with $\Delta$, consistent with a scaling coefficient $c_2 \sim 0.5$. In contrast, for NSTX \#139047, $\beta_p$ increases linearly across the pedestal, consistent with $c_2 \sim 1$. It should be highlighted that the local critical $\beta_p/\Delta$ is obtained using globally consistent equilibria.
 
\subsection{Pedestal evolution in 2\textsuperscript{nd} stability}

If local IBMs or KBMs are found to be stable, an alternative mechanism must exist to constrain the pedestal pressure gradient. While (slab) ITGs, ETGs and MTMs may be unstable in the pedestal, they primarily drive thermal transport and therefore do not necessarily limit the pressure gradient, which may continue to increase through density accumulation. TEMs can also be unstable in the pedestal region; however, their role as an effective pedestal constraint remains insufficiently understood. Global gyro-kinetic simulations performed for JET plasmas have demonstrated that global effects can destabilize KBMs in the 2\textsuperscript{nd} stability regime even when local KBMs are stable \cite{nonlocal}. Nevertheless, systematically investigating this mechanism using global gyro-kinetic simulations is impractical due to their prohibitive computational cost.
\begin{table}[t!]
\centering
\begin{tabular}{ | c | c | c | c | }
\hline
 & {\bf DIII-D} & {\bf NSTX} \\
\hline
pulse \# & 131999 & 139047 \\
\hline
$R_0$ [m] & 1.695 & 0.863 \\
\hline
$r$ [m] & 0.579 & 0.599 \\
\hline
$\kappa$ & 1.807 & 2.312 \\
\hline
$\delta$ & 0.148 & 0.536 \\
\hline
$B_T$ [T] & 2.098 & 0.486 \\
\hline
$I_p$ [MA] & 1.487 & 0.965 \\
\hline
$q_{95}$ & 3.615 & 8.849 \\
\hline
$\beta_N$ & 1.851 & 4.778  \\
\hline
$n_{e,ped}$ [m$^{-3}$] & 5.9$\cdot10^{19}$ & 4.8$\cdot10^{19}$ \\
\hline
$n_{e,ped}/n_{e,sep}$ & 4.0 & 1.6  \\
\hline
$n_{e,0}/n_{e,ped}$ & 1.5 &  1.5 \\
\hline
$T_{e,sep}$ [eV] & 75 & 80  \\
\hline
\end{tabular}
\caption{Table with global plasma parameters for DIII-D \#131999 and NSTX \#139047 plasmas.}
\label{specs}
\end{table}

An alternative approach is provided by the global ideal MHD model, which introduces destabilizing mechanisms through non-locality and the coupling of kink/peeling modes and finite $n$ effects on ballooning modes in the 2\textsuperscript{nd} stability regime \cite{localPB1,localPB2}. Within ideal MHD theory, it has been shown that global and finite $n$ effects can significantly modify the local infinite $n$ stability boundary, both when calculated to leading order in $1/n$ \cite{localPB1} and moreso in higher order calculations \cite{localPB2}. Although access to the 2\textsuperscript{nd} stability regime remains possible, the corresponding $\hat{s}$–$\alpha$ ballooning boundary is shifted toward lower values of $\hat{s}$ and higher values of $\alpha$, as illustrated schematically in Fig.\ref{glob}.

The ELITE code is employed to assess the stability of high $n$ modes. For toroidal mode numbers $n \sim (\rho^* q)^{-1}$, where $\rho^*=\rho_s/r$, the resulting global modes are nearly local in nature, as they are strongly localized within the steep-gradient region of the pedestal. In the present analysis, modes with $k_y \rho_s < 0.25$ are assumed to be non-local, since their radial extent may exceed the pedestal width, while modes with $k_y \rho_s > 0.5$ are expected to be stabilized by finite ion Larmor radius effects and are therefore excluded. This ordering is consistent with linear gyro-kinetic simulations, which typically find KBM growth rates to peak at relatively small values of $k_y \rho_s$. ELITE calculations are then used to determine the critical $\beta_{p,\mathrm{ped}}$ at which modes with $k_y \rho_s \simeq 0.25$–$0.5$ become marginally unstable. Notably, even at high $n$, the inclusion of the kink/peeling contribution is essential for instability, as the ballooning drive alone is insufficient to overcome the stabilizing field-line bending.
\begin{figure}[t!]
\centering
a) \includegraphics[height=5cm]{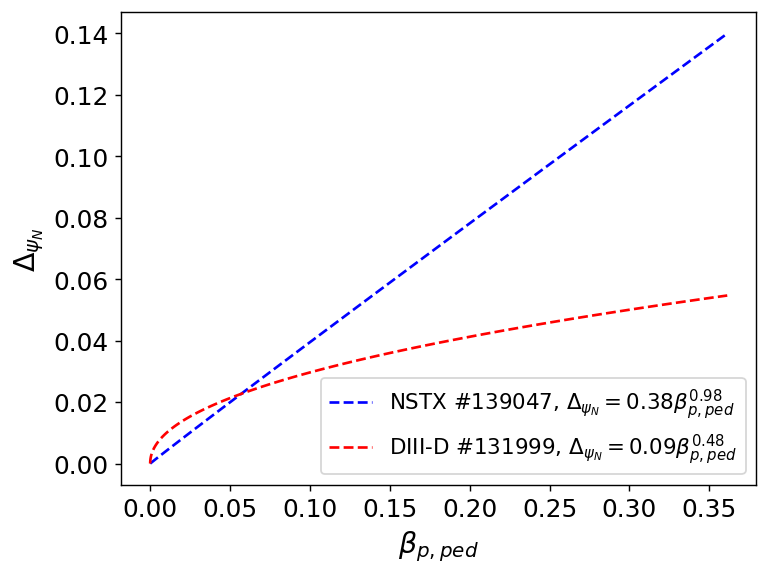} \\ b) \includegraphics[height=5cm]{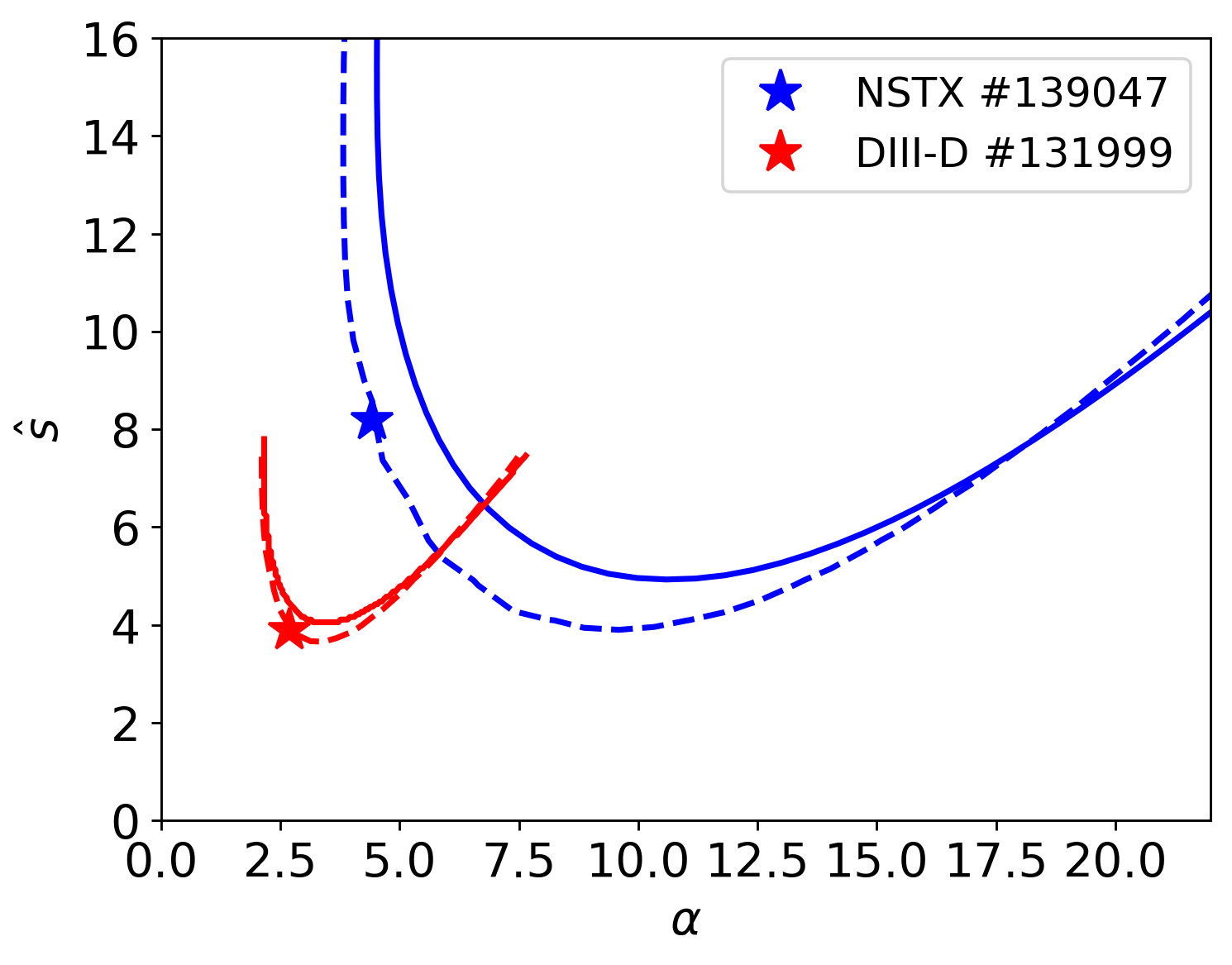}
\caption{a) Comparison of the pedestal width scaling with $\beta_{p,ped}$ between DIII-D \#131999 and NSTX \#139047 and b) $\hat{s}-\alpha$ kinetic ballooning stability boundary at the middle of the pedestal between DIII-D and NSTX plasmas. The solid lines represent the ideal ballooning stability boundary.}
\label{tokcomp}
\end{figure}
\begin{figure}[t!]
\centering
a) \ \includegraphics[height=5cm]{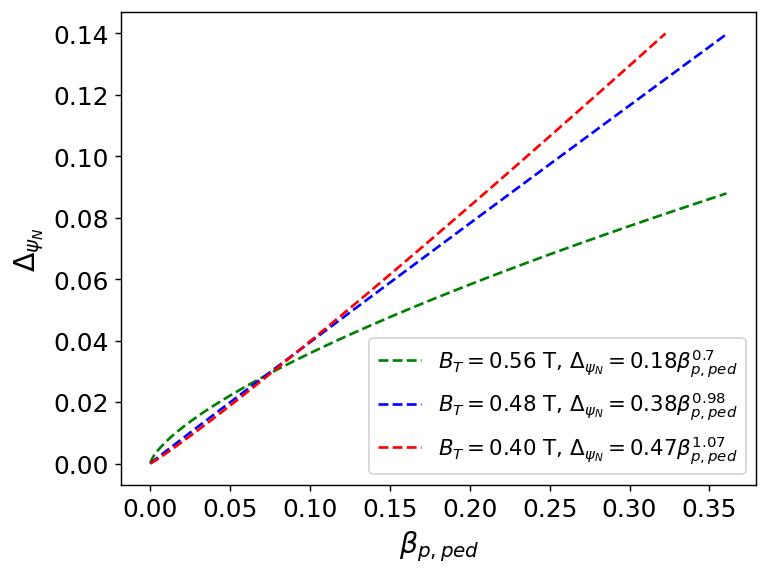} \\ b) \ \includegraphics[height=5cm]{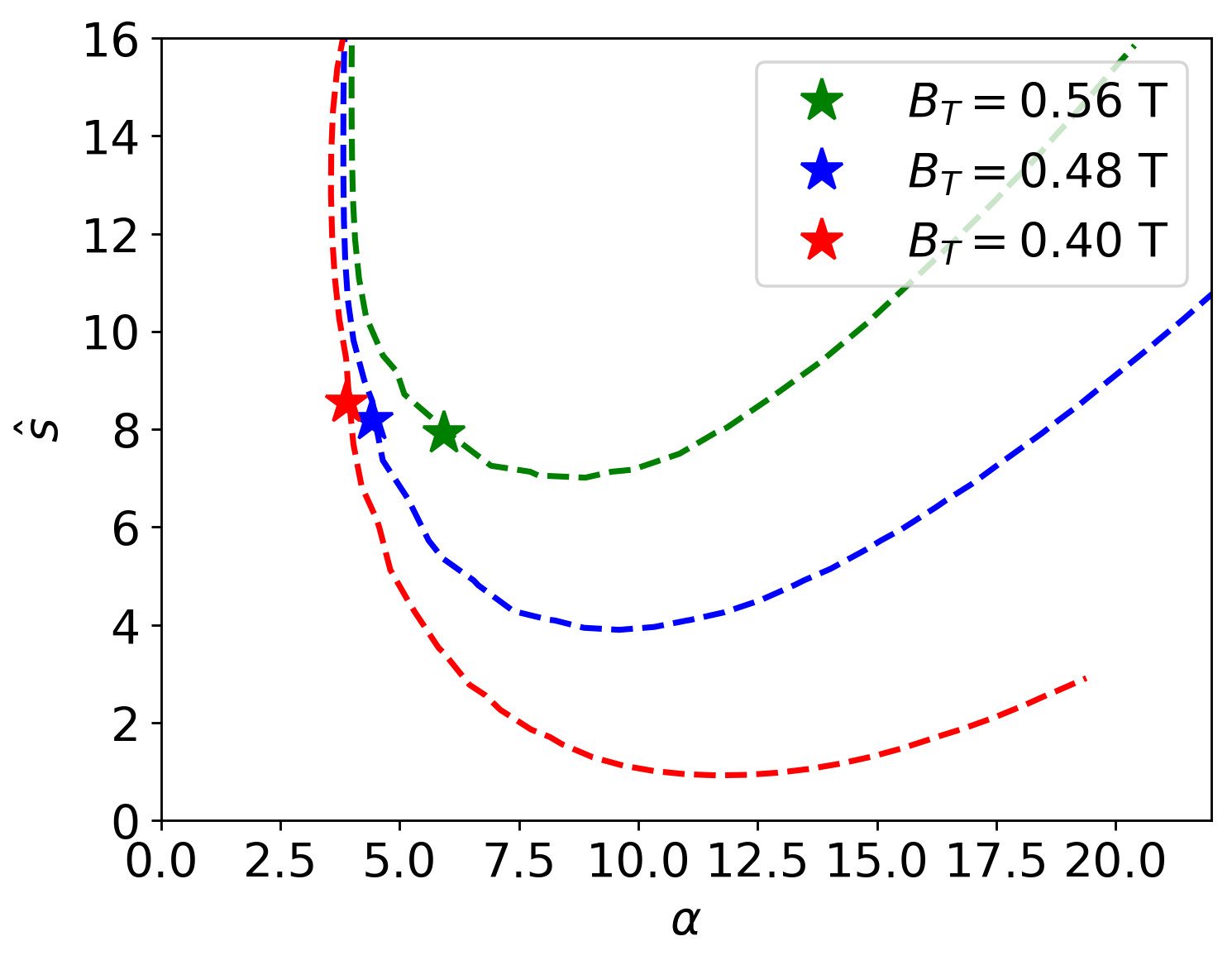}
\caption{a) Comparison of the pedestal width $\Delta_{\psi_N}$ scaling with $\beta_{p,ped}$ and b) the $\hat{s}-\alpha$ stability boundary at the middle of the pedestal considering different toroidal magnetic field $B_T$ on the geometric axis for NSTX \#139047.}
\label{field}
\end{figure}
\begin{figure}[t!]
\centering
a) \ \includegraphics[height=5cm]{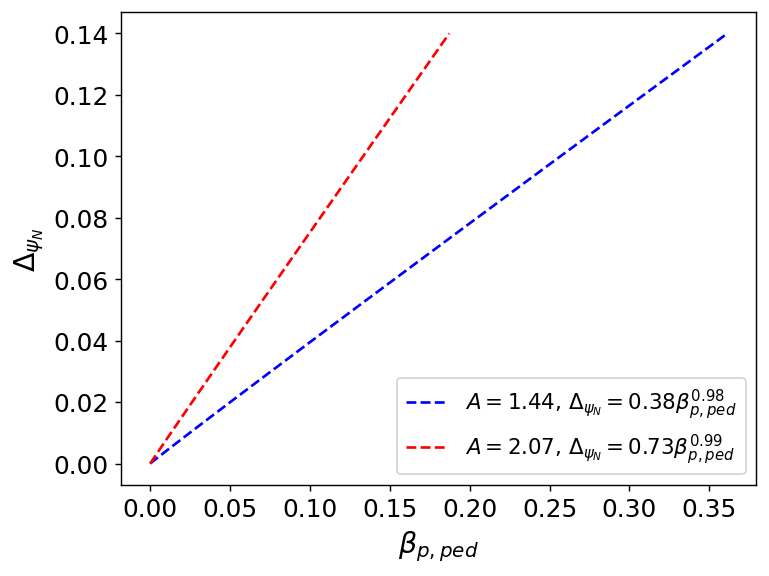} \\ b) \ \includegraphics[height=5cm]{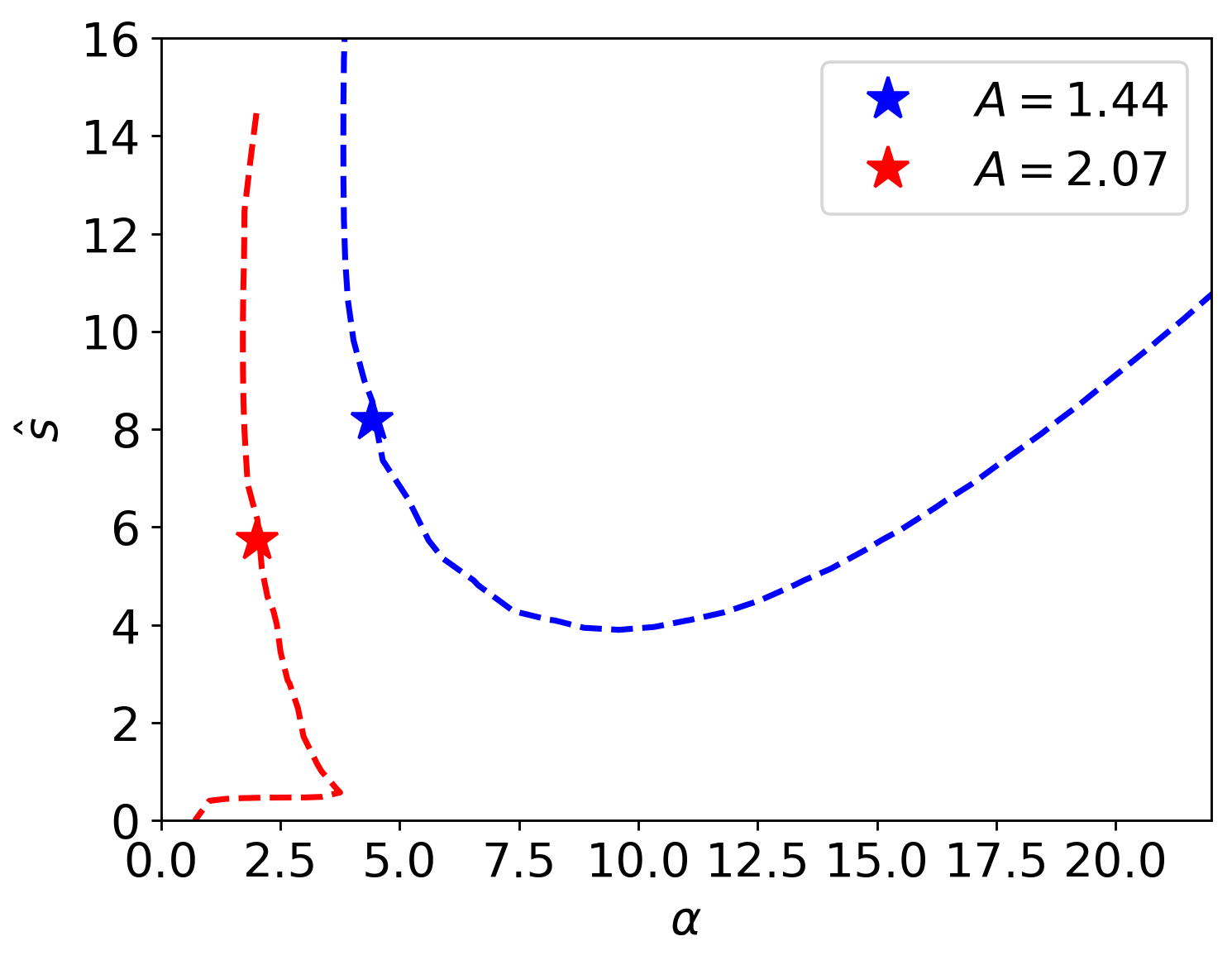} 
\caption{a) Comparison of the pedestal width $\Delta_{\psi_N}$ scaling with $\beta_{p,ped}$ and b) the $\hat{s}-\alpha$ stability boundary at the middle of the pedestal considering different aspect ratio $A$ for NSTX \#139047.}
\label{aspect}
\end{figure}
\begin{figure}[t!]
\centering
a) \ \includegraphics[height=5cm]{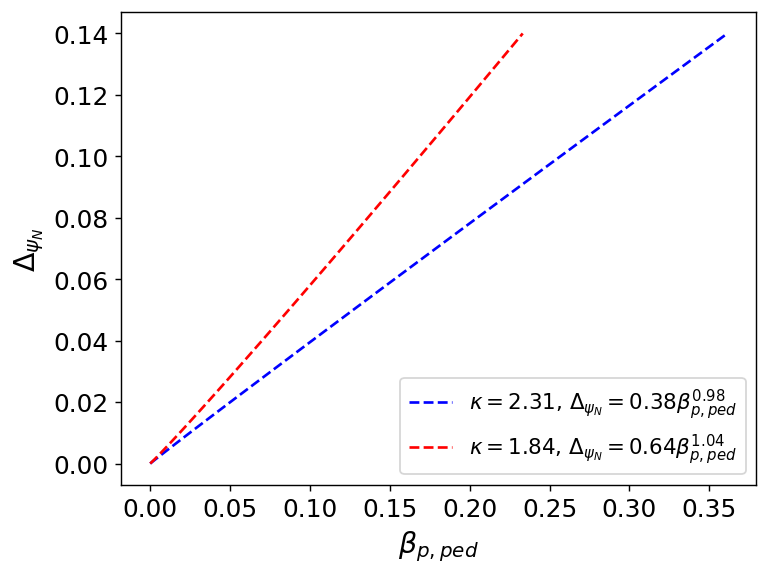} \\ b) \ \includegraphics[height=5cm]{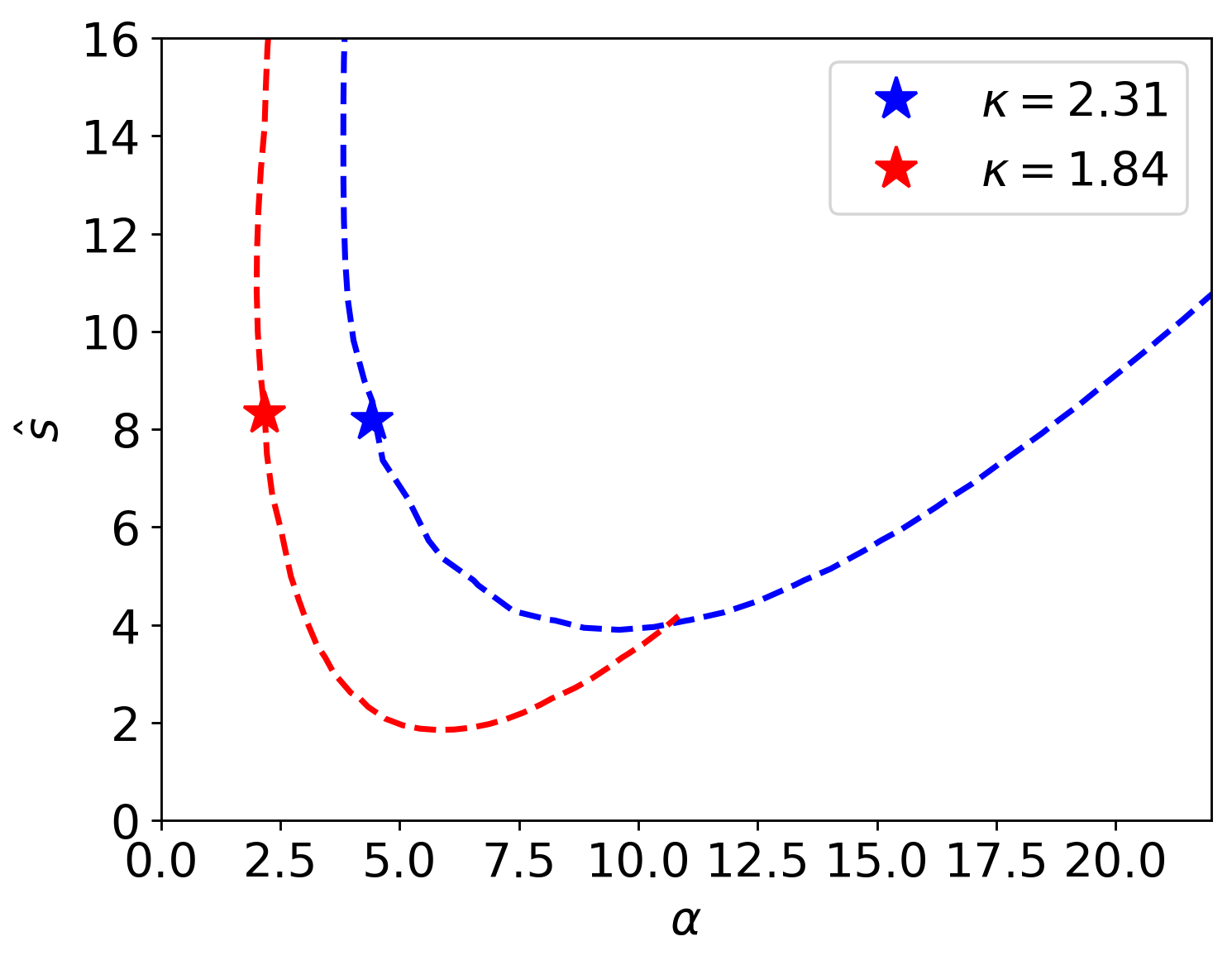} 
\caption{a) Comparison of the pedestal width $\Delta_{\psi_N}$ scaling with $\beta_{p,ped}$ and b) the $\hat{s}-\alpha$ stability boundary at the middle of the pedestal considering different elongation $\kappa$ for NSTX \#139047.}
\label{kappa}
\end{figure}
\begin{figure}[t!]
\centering
a) \ \includegraphics[height=5cm]{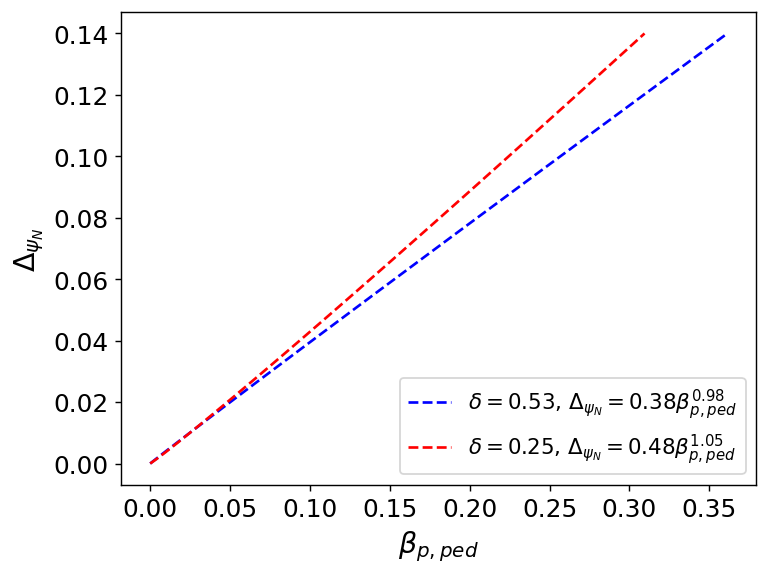} \\ b) \ \includegraphics[height=5cm]{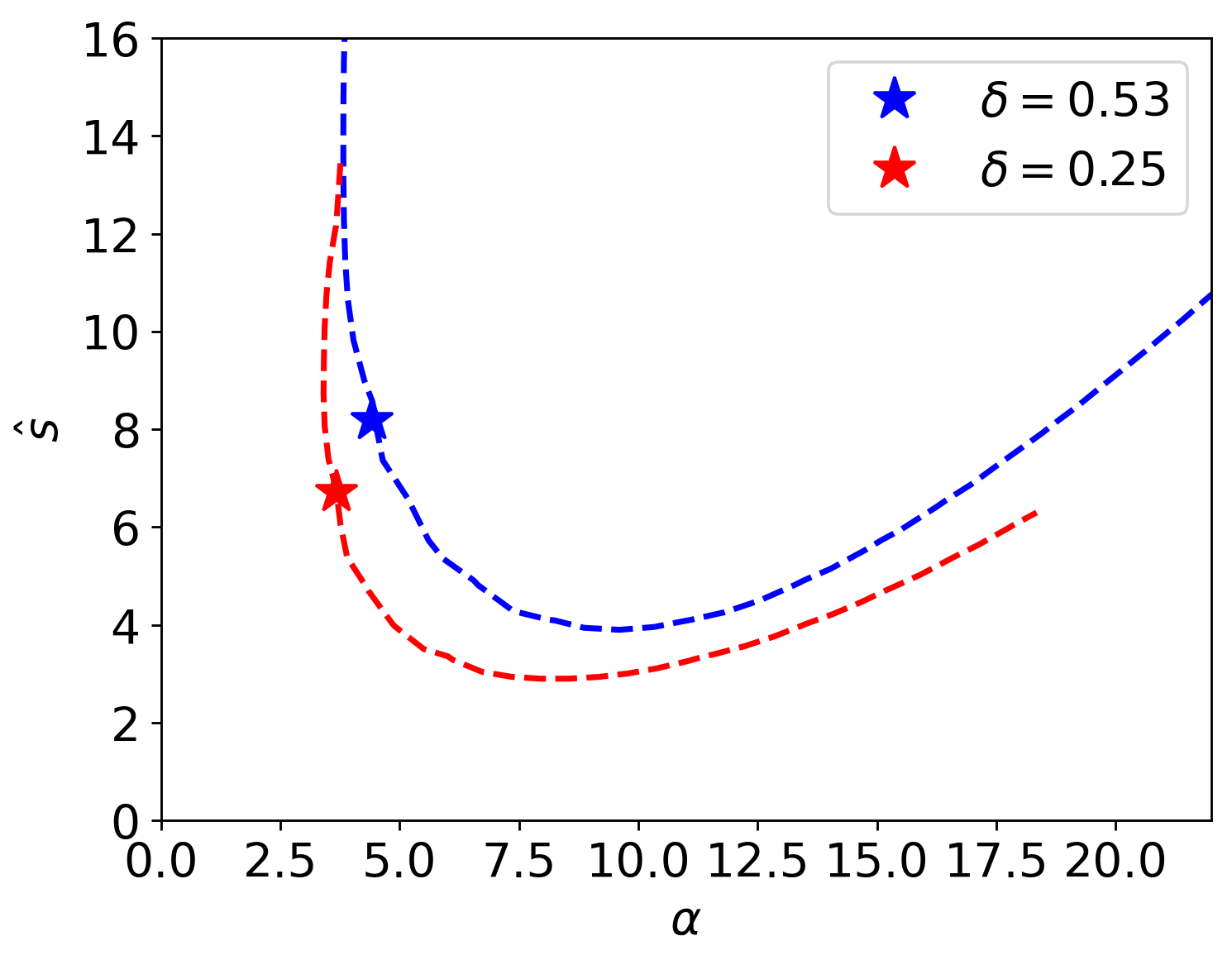} 
\caption{a) Comparison of the pedestal width $\Delta_{\psi_N}$ scaling with $\beta_{p,ped}$ and b) the $\hat{s}-\alpha$ stability boundary at the middle of the pedestal considering different triangularity $\delta$ for NSTX \#139047.}
\label{delta}
\end{figure}
\begin{figure}[t!]
\centering
a) \ \includegraphics[height=5cm]{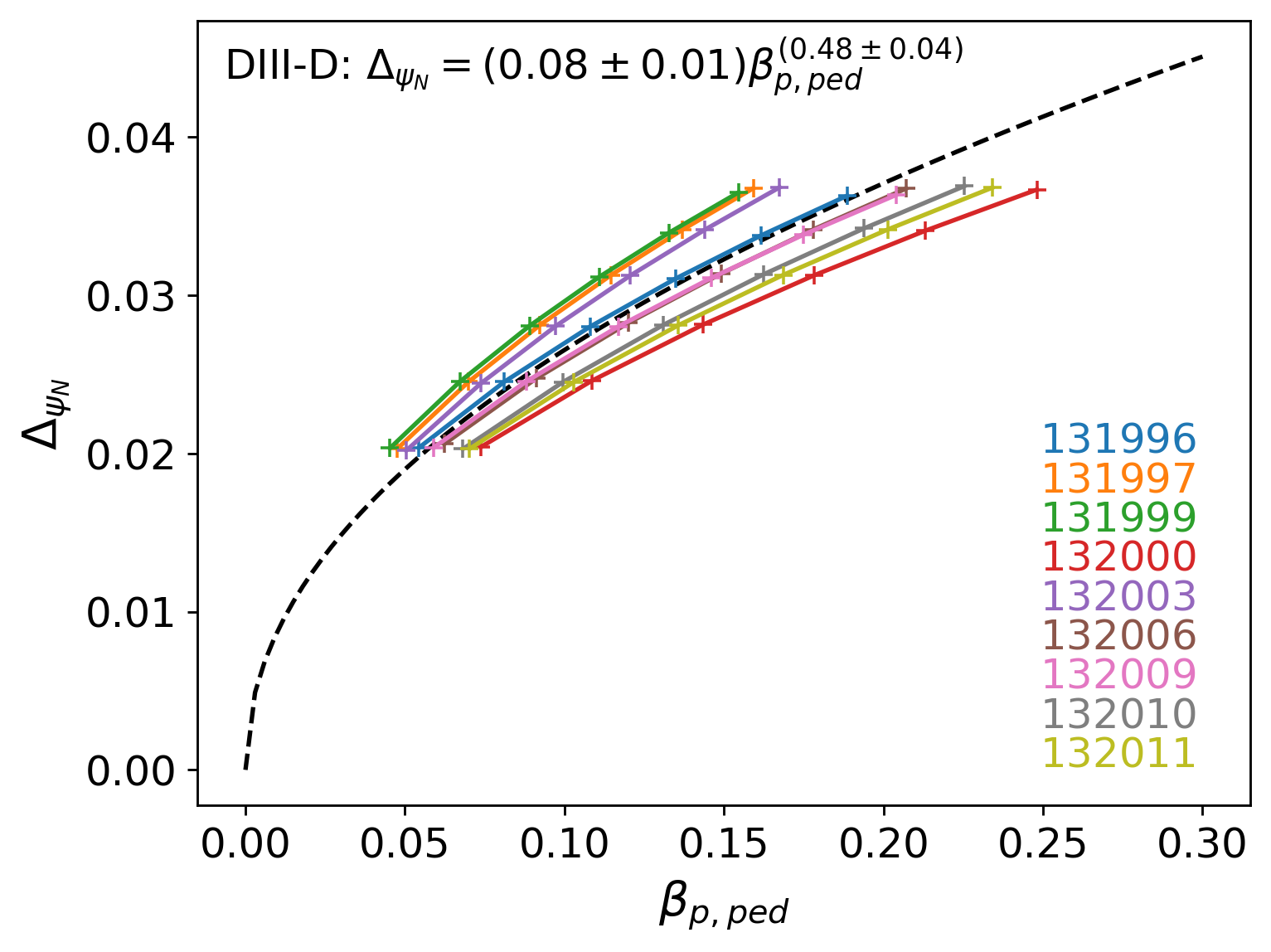} \\ b) \ \includegraphics[height=5cm]{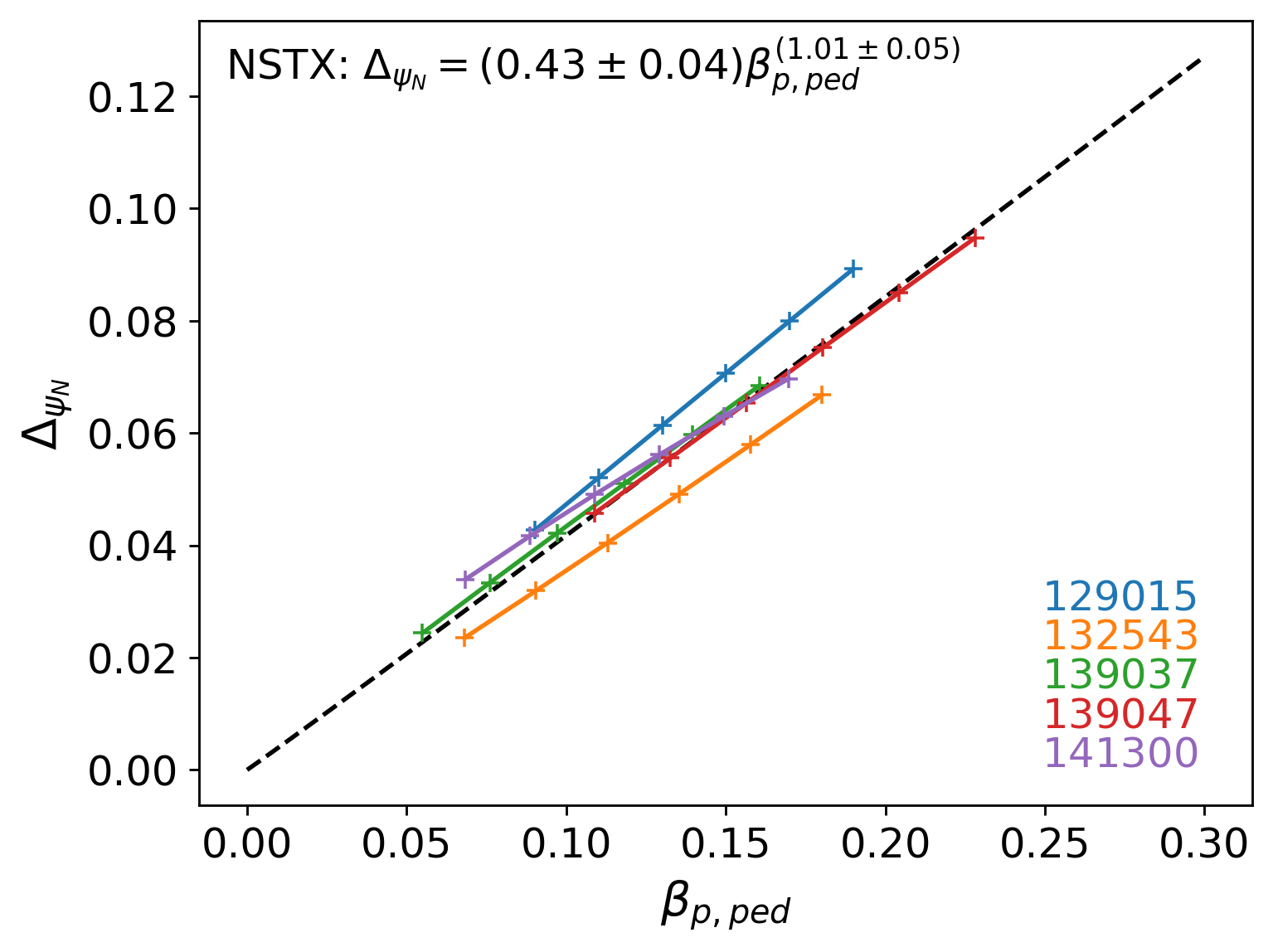} 
\caption{Pedestal width $\Delta_{\psi_N}$ and height $\beta_{p,ped}$ scaling over a wide range of plasma discharges from a) DIII-D and b) NSTX using the GFS code. The colored lines indicated the particular pulse while the black line indicates the average trend.}
\label{scale_stat}
\end{figure}
 
\section{Local pedestal width scaling}\label{sec3}

The application of GFS as a predictive tool for the pedestal width (as a fucntion of height) is assessed using ELMy H-mode plasmas from DIII-D and NSTX, with the ultimate goal of determining a scaling of the form $\Delta_{\psi_N} = c_1 \beta_{p,\mathrm{ped}}^{c_2}$.
The plasma parameters for the different tokamak configurations considered in this study are summarized in Table\ref{specs}. As shown in Fig.\ref{tokcomp}, the predicted $\Delta_{\psi_N} = c_1 \beta_{p,\mathrm{ped}}^{c_2}$ scalings for both machines are consistent with existing empirical scalings. For DIII-D discharge \#131999, the fit yields $c_1=0.09$ and $c_2=0.48$. For NSTX discharge \#139047, the predicted scaling is consistent with the results of Diallo \textit{et al.}\cite{scaleNSTX}, with $c_1=0.38$ and $c_2=0.98$. These results demonstrate the applicability of GFS for determining the KBM critical $\beta_{p,\mathrm{ped}}$ across a wide range of plasma parameters, in both conventional and low aspect-ratio tokamaks.

A pronounced difference in the pedestal width scaling is observed between NSTX and DIII-D. While DIII-D operates at conventional aspect ratio and both NSTX and MAST are low aspect-ratio devices, the observed deviation cannot be attributed solely to aspect ratio effects. Instead, the differences between DIII-D \#131999 and NSTX \#139047 arise from variations in toroidal magnetic field $B_T$, plasma current $I_p$, and plasma shaping parameters, including elongation $\kappa$ and triangularity $\delta$. These differences lead to distinct $q$-profiles and qualitative variations in pedestal stability, particularly with respect to the local $\hat{s}$–$\alpha$ diagram and access to the 2\textsuperscript{nd} stability regime, as illustrated in Fig.\ref{tokcomp}.

\begin{itemize}
\item{\bf Parametric dependence of NSTX \#139047 KBM stability}
\end{itemize}

To identify the main drivers responsible for the different pedestal width scalings observed in DIII-D and NSTX, the equilibrium parameters of the NSTX \#139047 plasma are varied individually. The impact of each parameter on the predicted scaling is then assessed. We begin by varying the toroidal magnetic field $B_T$ by $\pm15\%$. Increasing $B_T$ raises the $q$-profile, shifts the $\hat{s}$–$\alpha$ stability boundary upward, and facilitates access to the 2\textsuperscript{nd} stability regime, as shown in Fig.\ref{field}. As $B_T$ increases, the pedestal approaches the 2\textsuperscript{nd} stable region and a pronounced reduction in the scaling exponent $c_2$ is observed.

Geometric effects are also found to play a significant role in determining pedestal performance. The low aspect ratio $A$ of NSTX enhances pedestal performance, even though the pedestal remains limited by the first-stability boundary, as illustrated in Fig.\ref{aspect}. Increasing elongation $\kappa$ leads to a higher pedestal pressure $\beta_{p,\mathrm{ped}}$ for a fixed pedestal width $\Delta_{\psi_N}$, as shown in Fig.\ref{kappa}. Similarly, increased triangularity $\delta$ has a beneficial impact on $\beta_{p,\mathrm{ped}}$, as shown in Fig.\ref{delta}. Nevertheless, due to the large magnetic shear $\hat{s}$ characteristic of NSTX, the center of the pedestal remains far from the ``nose'' of the stability boundary even under strong shaping, resulting in $c_2 \sim 1$. It is noted that reducing plasma shaping and increasing the aspect-ratio, while keeping $B_T$ and $I_P$ fixed, leads to a reduction of the $q$-profile.

The parametric dependence of the pedestal width on geometric parameters obtained using GFS is in good agreement with the results of Ref\cite{parisi2}. Specifically, GFS predicts $\Delta_{\psi_N} \propto \kappa^{-1.86}$, $0.67^{\delta}$, and $A^{1.67}$, compared to $\kappa^{-1.8}$, $0.5^{\delta}$, and $A^{1.5}$ reported in Ref\cite{parisi2}.

Access to the 2\textsuperscript{nd}  stability regime—commonly observed in DIII-D pedestals—is found to correlate strongly with a reduced scaling exponent $c_2$. To interpret this behavior, we follow the arguments presented in Ref\cite{scaleDIIID}. The pedestal width may be approximated as $\Delta_{\psi_N} \sim {\beta_{p,\mathrm{ped}}}/{\alpha_{\mathrm{crit}}}$, where $\alpha_{\mathrm{crit}} \sim \mathrm{d}\beta_p/\mathrm{d}\Delta$ is the critical pressure gradient. Near the ``nose'' of the stability boundary, $\alpha_{\mathrm{crit}} \sim \hat{s}^{-1/2}$. In addition, in bootstrap-current-dominated pedestals the magnetic shear scales as $\hat{s} \sim 1/\langle j \rangle \sim 1/\beta_{p,\mathrm{ped}}$. Substituting these relations yields $\Delta_{\psi_N} \propto \beta_{p,\mathrm{ped}}^{1/2}$. In contrast, at high magnetic shear where $\alpha_{\mathrm{crit}}$ is approximately constant, the pedestal width scales linearly with pedestal pressure, leading to $\Delta_{\psi_N} \propto \beta_{p,\mathrm{ped}}$.

These results demonstrate that proximity to, and access into, the 2\textsuperscript{nd} stability regime are central to understanding the observed differences in pedestal width scaling between NSTX and DIII-D. Finally, it is worth noting that at low shaping, access to 2\textsuperscript{nd}  stability may be inhibited by edge-localized kink/peeling modes, which can reduce $\alpha_{\mathrm{crit}}$ as $\hat{s}$ decreases. However, such an equilibrium evolution is not physically self-consistent, owing to the strong coupling between magnetic shear, bootstrap current, and pressure gradient, that does not allow $\alpha$ and $\hat{s}$ to decrease simultaneously.

\begin{itemize}
\item{\bf Statistical fit of $\bm{ \Delta_{\psi_N}=c_1\beta_{p,\mathrm{ped}}^{c_2} }$ scaling on DIII-D and NSTX plasmas}
\end{itemize}

Finally, using a set of representative ELMy H-mode plasmas from DIII-D and NSTX, GFS is applied to compute the pedestal width scaling. As shown in Fig.\ref{scale_stat}, excellent statistical agreement is obtained for both devices. For DIII-D, the scaling is $\Delta_{\psi_N} = 0.08\,\beta_{p,\mathrm{ped}}^{0.48}$, while for NSTX it is $\Delta_{\psi_N} = 0.43\,\beta_{p,\mathrm{ped}}^{1.01}$. These results are consistent with experimentally inferred scalings for conventional and low aspect-ratio tokamaks.

A clear qualitative distinction is observed between the two datasets. In all DIII-D cases considered, the pedestals are found to operate close to the ``nose'' of the $\hat{s}$–$\alpha$ stability boundary, whereas the NSTX pedestals are limited by the 1\textsuperscript{st} ballooning boundary at high magnetic shear, where the critical pressure gradient $\alpha_{\mathrm{crit}}$ is approximately constant. This fundamental difference in $\hat{s}$–$\alpha$ stability behavior is identified as the primary physical mechanism underlying the contrasting pedestal width scalings in DIII-D and NSTX.

\section{Global effects on local 2\textsuperscript{nd} stable pedestals}\label{sec4}

It becomes apparent from Section~\ref{sec3} that, under certain conditions, DIII-D pedestals can access the local 2\textsuperscript{nd} stable region. Indeed, in several experimental cases plasma parameters exist for which the pedestal is locally ideal-ballooning 2\textsuperscript{nd} stable. In such situations, the approach described above cannot provide a well-defined critical value of $\mathrm{d}\beta_p/\mathrm{d}\Delta$, and additional physical effects must be taken into account. Because fully global gyro-kinetic (or 6D kinetic, which avoid the locality inherent in the gyro-kinetic ordering) calculations are generally impractical for predictive pedestal studies, reduced fluid and MHD-based models become an attractive alternative. Accordingly, the ELITE code is employed to examine the marginal stability of nearly local, high $n$, ideal MHD modes.

\begin{figure}[t!]
\centering
\includegraphics[height=5cm]{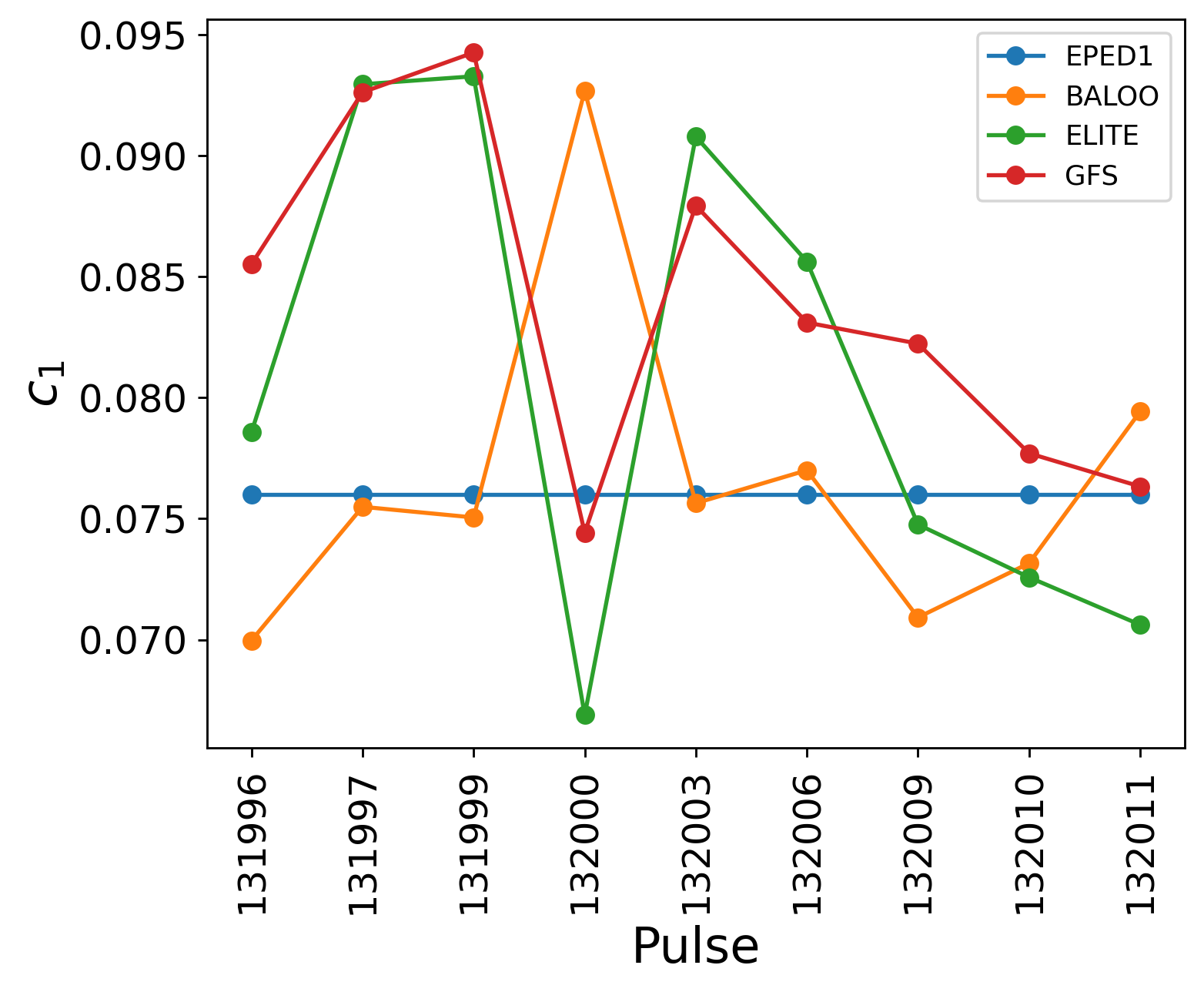}
\caption{Comparison of the $c_1$ coefficient for $\Delta_{\psi_N}\propto\beta^{0.5}_{p,ped}$ scaling between BALOO, ELITE and GFS using a set of DIII-D plasma discharges.}
\label{gfs_elite_bal}
\end{figure}

To illustrate this point, a set of DIII-D discharges is considered that are locally KBM-limited according to GFS calculations, but exhibit partial or complete access to $n=\infty$ ideal-ballooning 2\textsuperscript{nd} stability as determined by BALOO \cite{baloo}. These cases are used to compare the coefficient $c_1$ obtained from ELITE limited pedestals, using $k_y\rho_s=0.25$, with those obtained from GFS and BALOO limited pedestals. In this exercise, the exponent of $\beta_{p,\mathrm{ped}}$ is fixed at $c_2=0.5$, consistent with the scaling reproduced using GFS. As shown in Fig.\ref{gfs_elite_bal}, the GFS and ELITE results exhibit good qualitative and quantitative agreement, whereas the scaling obtained using BALOO deviates even qualitatively from the kinetic and finite $n$ results. This discrepancy has important implications for EPED predictions, since the pedestal width scaling strongly influences the maximum achievable pedestal pressure.

The impact of access to 2\textsuperscript{nd} stability on the accuracy of the BCP technique is particularly evident in DIII-D \#132000. In this discharge, the entire central half of the pedestal is found to be 2\textsuperscript{nd} stable according to the $n=\infty$ IBM calculation, meaning that the BCP technique has no first stable region to anchor the calculation.  If the BCP technique is nonetheless employed, it identifies the pedestal as 2\textsuperscript{nd} stable even at relatively low pedestal pressure, leading to an artificially large value of $c_1$. In contrast, both GFS and ELITE predict a well-defined ballooning stability boundary at significantly higher pedestal pressure, arising from kinetic and finite $n$ effects, respectively, which in turn yields a substantially smaller value of $c_1$.

\section{Kinetic and finite $n$ effects on EPED predictions}\label{sec5}

To further demonstrate the applicability of these approaches, the DIII-D pedestal cases considered in this study are analyzed using EPED with GFS, ELITE, and BALOO employed as the KBM constraint. As a reminder, in the majority of these pedestals local ideal stability indicates that a significant fraction of the pedestal has access to the 2\textsuperscript{nd} stable region, with the stability boundary ultimately set by kinetic or finite $n$ effects. 

\begin{figure}[t!]
\centering
a) \includegraphics[height=5cm]{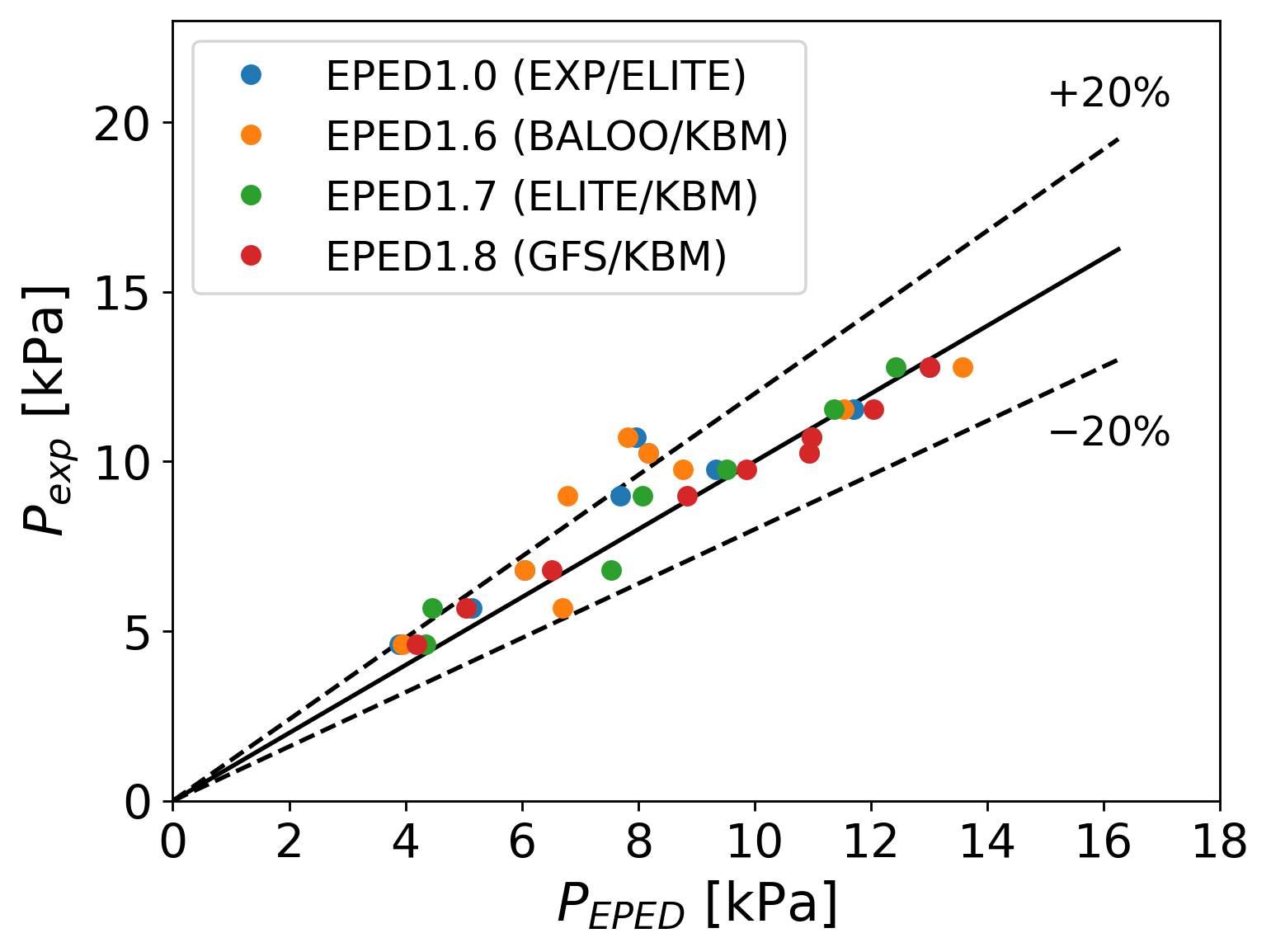} \\ b) \includegraphics[height=5cm]{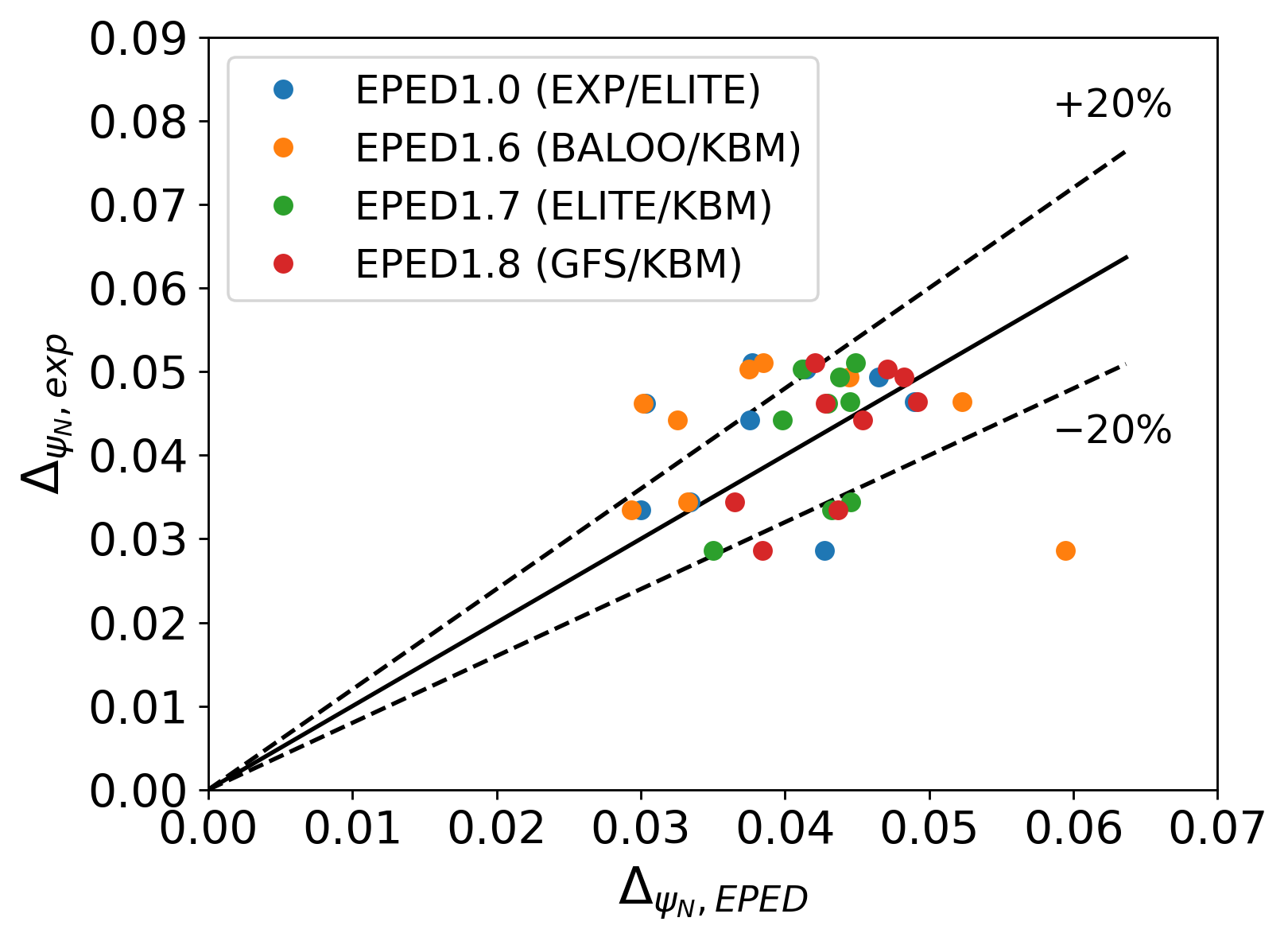} 
\caption{Comparison of a) the pedestal pressure ($P=2n_eT_e$) and b) the width $\Delta_{\psi_N}$ between EPED1.0 (EXP/ELITE), EPED1.6 (BALOO/ELITE), EPED1.7 (ELITE/ELITE) and EPED1.8 (ELITE/GFS) with DIII-D experimental data.}
\label{eped_comp}
\end{figure}

For each plasma discharge, GFS, ELITE, and BALOO are used to determine the pedestal width scaling, which is then supplied to EPED to compute the global peeling--ballooning stability boundary using ELITE. In addition, the standard EPED1.0 scaling, $\Delta_{\psi_N}=0.076\,\beta_{p,\mathrm{ped}}^{0.5}$, is included for comparison. As shown in Fig.\ref{eped_comp}, improved agreement with experimental measurements is obtained for both the pedestal pressure, $P_{\mathrm{ped}}=2n_{e,\mathrm{ped}}T_{e,\mathrm{ped}}$, and the pedestal width, $\Delta_{\psi_N}$, when kinetic or finite $n$ width scalings are used. 

Quantitatively, the mean error of EPED1.0 (EXP/ELITE) is $\epsilon_P=0.12\%$ and $\epsilon_\Delta=0.18\%$, while for EPED1.6 (BALOO/ELITE) it is $\epsilon_P=0.15\%$ and $\epsilon_\Delta=0.28\%$ (note that the EPED1.6 error is high in part because it is being applied to cases with broad regions of 2nd stability where its use is not recommended). Comparatively, EPED1.7 (ELITE/ELITE) yields $\epsilon_P=0.07\%$ and $\epsilon_\Delta=0.16\%$, and EPED1.8 (GFS/ELITE) further improves the agreement with $\epsilon_P=0.05\%$ and $\epsilon_\Delta=0.13\%$. 

These results demonstrate that employing GFS or ELITE to determine the KBM constraint leads to significantly improved agreement with experimental pedestal pressure and width, underscoring the importance of kinetic and finite $n$ effects in pedestal regimes close to 2\textsuperscript{nd} stability. These two effects, however, should not be regarded as equivalent, as they capture distinct physical mechanisms. As discussed earlier, a simultaneous treatment of both effects is not currently feasible. Consequently, the two approaches are complementary and together provide an uncertainty band for pedestal predictions. The inclusion of global finite $n$ effects becomes particularly important in regimes where both IBMs and KBMs are 2\textsuperscript{nd} stable.

\section{Summary}\label{sec6}

This work focused on calculating the KBM-critical pedestal width $\Delta_{\psi_N}$ as a function of height $\beta_{p,\mathrm{ped}}$ considering that it typically takes the form of $\Delta_{\psi_N} = c_1\beta_{p,\mathrm{ped}}^{c_2}$. The required improvements are motivated by the unique characteristics of NSTX pedestals, that are typically not described by the local IBM stability, and the recent resolution of the discrepancy attributed to kinetic effects. The development of the GFS code allowed for an accurate and fast numerical algorithm to perform KBM stability calculations as shown through this work. 

Predictions in good agreement with observation are reproduced considering the KBM critical $\beta_{p,\mathrm{ped}}$ as calculated from GFS for a number of DIII-D and NSTX plasmas. A comparison between DIII-D and NSTX shows that the DIII-D pedestals are typically close to the ``nose'' of the $\hat{s}-\alpha$ stability diagram, while NSTX pedestals are typically found at higher $\hat{s}$ in the 1\textsuperscript{st} stability boundary for the cases under consideration. Through changes on the plasma parameters and geometry of NSTX \#139047, the originally calculated linear width scaling approached the square root dependence when the pedestal approached the ``nose'' of the stability diagram.

When the pedestal gets access to the local IBM or KBM 2\textsuperscript{nd} stability region, a transport mechanism should exist that constrains the evolution of the pressure gradient. Global equilibrium and finite $n$ effects have been identified to provide a mechanism for the destabilization of ballooning modes for local 2\textsuperscript{nd} stable pedestals. In order to capture such effects, ELITE is used as a non-local KBM proxy for capturing unstable $k_y \rho_s\sim0.25-0.5$ modes localized within the pedestal region to compute a critical $\beta_{p,\mathrm{ped}}$. The coupling of kink/peeling and ballooning drive is essential and results in unstable modes even at high $n$ when local ballooning modes are 2\textsuperscript{nd} stable. 

Considering a set of DIII-D pedestals where partial or complete local IBM 2\textsuperscript{nd} stable access is observed, the predicted electron pedestal pressure and width as computed from EPED using the GFS or ELITE scaling are in good agreement with the experimental measurements, providing a systematic approach for computing the pedestal scaling when local IBMs are ideal 2\textsuperscript{nd} stable. It is highlighted that kinetic and finite $n$ effects, although both further destabilize the ballooning mode, are different in nature. 

To summarize, this work presents a set of complementary approaches that incorporate higher-fidelity physics into the calculated KBM constraints of EPED-like models. Kinetic effects become particularly important near marginal stability, especially in the vicinity of the “nose” of the ballooning stability boundary. When the pedestal accesses 2\textsuperscript{nd} stability for local ideal $n=\infty$ ballooning modes, kinetic effects can often provide sufficient destabilization for the operating point to intersect the kinetic ballooning boundary. In scenarios where both ideal and kinetic local ballooning modes remain 2\textsuperscript{nd} stable, global finite $n$ effects become critical for accurate calculation of the observed constraint on the pedestal gradient.

\section*{Acknowledgments}
This material is based upon work supported by the U.S. Department of Energy, Office of Science, Office of Fusion Energy Sciences, Theory Program and the NSTX Research Program, using the NSTX Fusion Facility, a DOE Office of Science user facility. We thank the NSTX experimental team for providing the data analyses.

DISCLAIMER: This report was prepared as an account of work sponsored by an agency of the United States Government. Neither the United States Government nor any agency thereof, nor any of their employees, makes any warranty, express or implied, or assumes any legal liability or responsibility for the accuracy, completeness, or usefulness of any information, apparatus, product, or process disclosed, or represents that its use would not infringe privately owned rights. Reference herein to any specific commercial product, process, or service by trade name, trademark, manufacturer, or otherwise, does not necessarily constitute or imply its endorsement, recommendation, or favoring by the United States Government or any agency thereof. The views and opinions of authors expressed herein do not necessarily state or reflect those of the United States Government or any agency thereof.

Notice: This manuscript has been authored by UT-Battelle, LLC, under contract DE-AC05-00OR22725 with the US Department of Energy (DOE). The US government retains and the publisher, by accepting the article for publication, acknowledges that the US government retains a nonexclusive, paid-up, irrevocable, worldwide license to publish or reproduce the published form of this manuscript, or allow others to do so, for US government purposes. DOE will provide public access to these results of federally sponsored research in accordance with the DOE Public Access Plan (https://www.energy.gov/doe-public-access-plan).

The data that support the findings of this study are available from the corresponding author upon reasonable request.

\end{document}